%% file: pres.tex
\documentclass[12pt,preprint]{aastex}
\usepackage{graphicx}
\usepackage{amssymb}
\usepackage{mathrsfs}

\newcommand{\bq}{\begin{equation}} 
\newcommand{\eq}{\end{equation}}

\shorttitle{A New Method to Calibrate the Magnitudes of Type Ia
Supernovae at Maximum Light}

\shortauthors{J.~L. Prieto, A. Rest, \& N.~B. Suntzeff}

\begin{document}

\title{A New Method to Calibrate the Magnitudes of Type Ia Supernovae at Maximum Light}

\author{Jos\'e Luis Prieto}
\affil{Department of Astronomy, The Ohio State University, 4055 McPherson Laboratory, 140 W. 18th Ave., Columbus, Ohio 43210}
\email{prieto@astronomy.ohio-state.edu}
\author{Armin Rest\altaffilmark{1} \& Nicholas B. Suntzeff}
\affil{Cerro Tololo Inter-American Observatory, National Optical Astronomy Observatory \altaffilmark{2}, Casilla 603, La Serena, Chile}
\email{rest@ctio.noao.edu, nsuntzeff@noao.edu}

\altaffiltext{1}{Goldberg Fellow.}
\altaffiltext{2}{The National Optical Astronomy Observatory is operated by the Association of Universities for Research in Astronomy, Inc., under cooperative agreement with the National Science Foundation.}

\begin{abstract}

We present a new empirical method for fitting multicolor light curves
of Type Ia supernovae.  Our method combines elements from two widely
used techniques in the literature: the \mbox{$\Delta$m$_{15}$}
template fitting method \citep{Phillips:1999} and the Multicolor Light-Curve 
Shape method \citep[MLCS;][]{Riess:1996}.  
An advantage of our technique is the ease of adding new colors, templates, or parameters 
to the fitting procedure. We use a large sample of published light curves to calibrate the
relations between the absolute magnitudes at maximum and the
post-maximum decline rate $\Delta {\rm m}_{15}$ in $BVRI$ filters. If
we perform a cut in reddening and compare an unreddened with a
reddened sample, we find that the two samples produce relations which are
marginally consistent with each other.  We find that individual
subsamples from a given survey or publication have significantly
tighter relationships between light curve shape and luminosity than
the relationship derived from the sum of all the samples, pointing to
uncorrected systematic errors in the photometry, mainly in $BV$
filters. Using our method, we calculate luminosity distances and host
galaxy reddening to 89 SNe in the Hubble flow and construct a low-$z$
Hubble diagram. The dispersion of the SNe in the Hubble diagram is
$\sigma=0.20$ mag, or an error of $\sim 9\%$ in distance to a single
SN.  Our technique produces similar or smaller dispersion in the
low-$z$ Hubble diagram than other techniques in the literature.

\end{abstract}


\section{Introduction}

Type Ia supernovae (Type Ia SNe), the thermonuclear explosions of 
accreting white dwarfs \citep{Arnett:1982}, are the most precise
distance indicators at cosmological distances.  Even though they are
not standard candles \citep{Pskovskii:1977,Branch:1987,Phillips:1993},
their absolute magnitudes at maximum light are closely correlated with
the shape of their light curves \citep{Phillips:1993}.  As precise
{\em standardizable} candles they have been used to trace the
expansion of the Universe as a function of redshift
\citep[see][]{Leib:2000}, leading to the discovery of the acceleration
of the expansion \citep{Riess:1998,Perlmutter:1999}, and the
transition from deceleration at early times to acceleration in the
present \citep{Tonry:2003,Riess:2004}.

All the information of the luminosity distance of a Type Ia SN is in
the multicolor light curves.  Different empirical methods have been
presented in the literature to fit Type Ia SNe light curves (LCs).
Among the most developed are: \mbox{$\Delta$m$_{15}$}, MLCS, and the
``stretch'' method.

The \mbox{$\Delta$m$_{15}$} method \citep{Hamuy:1996a,
Phillips:1999} uses six $BVI$ templates of nearby SNe 
\citep{Hamuy:1996d} covering a wide range of
light curve shapes which are parametrized by a single discrete value:
the 15 day post-maximum decline in magnitude in the $B$ band
\mbox{$\Delta$m$_{15}(B)$}. \citet{Germany:2004} provided a modified
version of the method by including the $R$ band in the fits, new
estimations of the $K$-corrections, and new templates from
well-observed nearby SNe.

In the MLCS method \citep{Riess:1996,Riess:1998} a continuous
parameter characterizes different light curve shapes. This parameter
is the difference between the absolute magnitude at maximum of a SN
and a fiducial value, constructed from a set of well observed SNe. 
The latest version of the method \citep{Jha:2006b} has added the
$U$ filter to $BVRI$ and better extinction estimators.

The stretch method \citep{Perlmutter:1997,Goldhaber:2001} measures the
light curve shape by simply adjusting the scale on the time axis by a
multiplicative factor.  This is an elegant solution, since the stretch
of the light curve is the combination of the cosmological redshift
effect (\mbox{$1+z$} factor) and the intrinsic light curve
shape. Different LCs are stretched in the time domain to fit a
template \citep{Leib:1988}. This method has been used mainly in $UBV$
filters because of the homologous nature of the light curves across
all light curve shapes. \mbox{$\Delta$m$_{15}$} and MLCS are more
robust to non-homologous variations as a function of light curve shape
in the redder filters.

Other methods have been proposed \citep{Tripp:1998, Tripp:1999,
Tonry:2003, Wang:2003, Wang:2005}. \citet{Wang:2005} found a tight
correlation between peak luminosities and the intrinsic colors at
$\sim$12 days after $B$ maximum. Using this correlation but with a
small sample of SNe, they obtain very low-dispersion Hubble diagrams.

The new technique presented in this paper is a combination of 
\mbox{$\Delta$m$_{15}$} and MLCS methods. We implement what we
consider the advantages from each method: the direct use of actual
(not interpolated) SNe LCs templates characterized by different
\mbox{$\Delta$m$_{15}$} values in the fitting method, and a proper
statistical model of the errors in the fitting parameters. We use a
simple mathematical framework to account for the non-uniform
distribution in the values of the \mbox{$\Delta$m$_{15}$}
parameter. Our method allows us to calculate a covariance matrix for
the fits as in MLCS. It also allows us to trivially add any new
supernova light curve or any new color as a template in $\Delta {\rm
m}_{15}$ without having to perform a training set calculation.

This paper is organized as follows: in Section 2 we present the
template set, composed by 14 templates with different decline rates;
in Section 3 we explain the mathematical technique to interpolate
between templates and the model to fit the multicolor light curves of
SNe; in Section 4 we obtain linear relations between the absolute
magnitudes at maximum and $\Delta {\rm m}_{15}$ in rest frame $BVRI$
filters studying a large sample of SNe; in Section 5 we construct a
Hubble diagram with a large sample of low-$z$ Type Ia SNe in the
Hubble flow, and compare our method with other results published in
the literature; and finally, the general conclusions of the results
obtained in the paper are presented in Section 6.

\section{Light curve templates}
\label{sec:1}

The template set, listed in Table \ref{table_1} is composed of the 6
$BVI$ templates of \citet{Hamuy:1996d} augmented with the $R$ filter
templates of \citet{Hamuy:1996d}, and 8 new $BVRI$ templates
constructed from published and unpublished data of well observed
nearby Type Ia SNe.  The SNe were selected to cover a wide range in
observed light curve shapes.  We recalculated the $\Delta {\rm
m}_{15}$ from a the spline interpolation of the $B$-band data.

We used the same recipe of \citet{Hamuy:1996d} to construct the LC
templates. A cubic spline interpolation was applied to the data
obtaining the time and magnitude at maximum in each filter, with
typical uncertainties of $\sim 0.2-0.5$ days and $\sim 0.02-0.05$ mag,
respectively.  The time axis in all the filters was parametrized by
the rest frame time relative to $B$ maximum.  In the cases where the
redshift in the Cosmic Mircowave Background (CMB) frame is
$z_{CMB}\geq 0.01$, the time axis was corrected for cosmological time
dilation dividing by \mbox{$1+z$}, and the magnitudes were
$K$-corrected to the rest frame.

We calculated the $K$-corrections in $BVRI$ filters using the recipe
of \citet{Hamuy:1993} and a set of spectrophotometry from different
epochs and SNe: \objectname{SN~1972E}, \objectname{SN~1990N},
\objectname{SN~1991T}, \objectname{SN~1991bg}, \objectname{SN~1992A},
and \objectname{SN~1994D}. We fitted the results with a cubic spline
to provide values of $K$-corrections in different epochs, with a
typical error $\sim 0.02$ mag.

Finally, the magnitudes at $B$ maximum were subtracted in all the
filters to have $B=V=R=I=0.0$ at \mbox{t$_{0}(B)$}.  The final
templates have interpolated rest frame magnitudes in the range $-5
\leq {\rm t-t}_{0}(B){\rm [days]} \leq 80$, with a sampling of one
day.
         
\begin{figure}
\includegraphics[angle=0,scale=0.8]{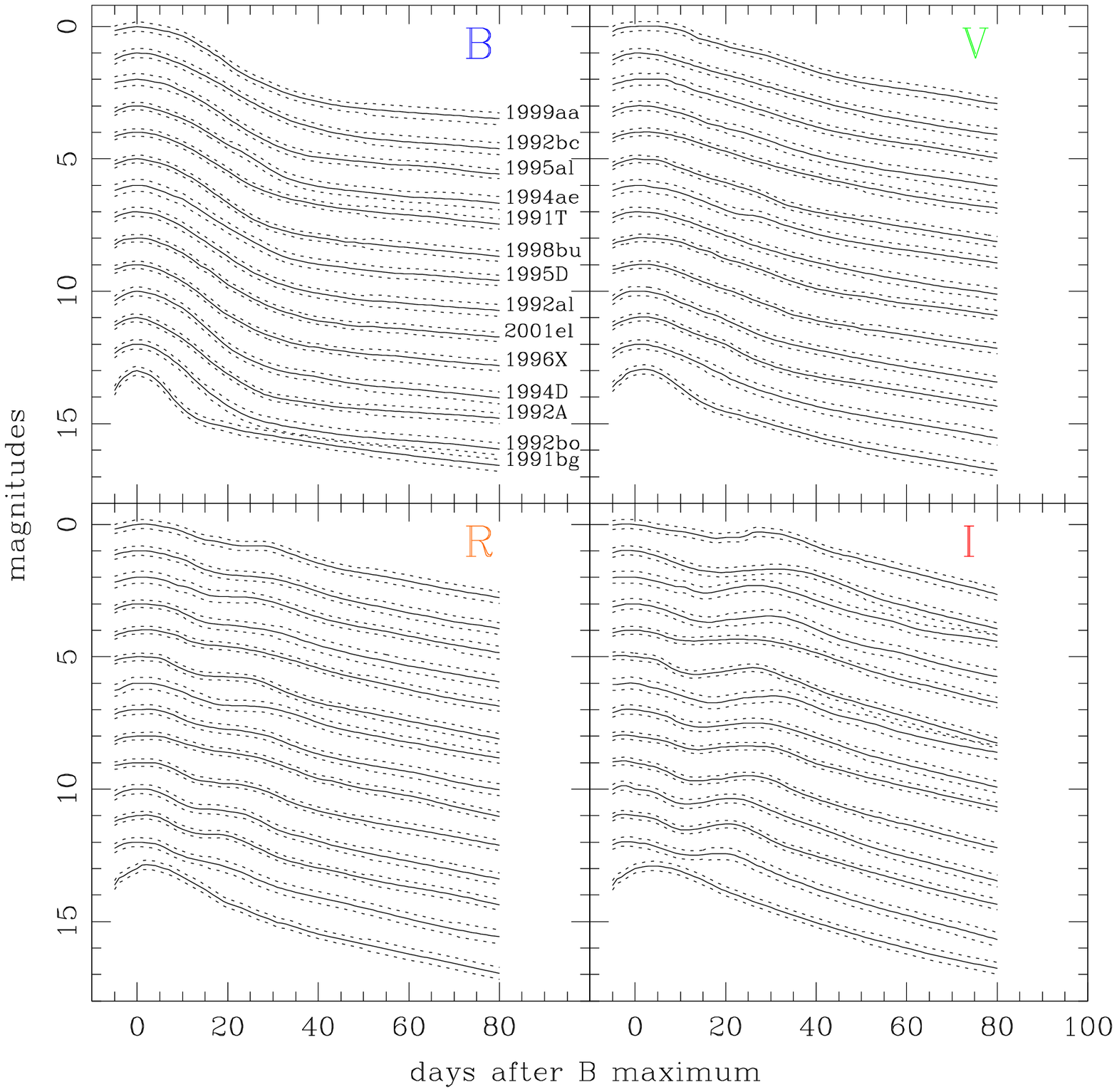}
\caption{Light curve template set in $BVRI$ filters. These 14
templates, 6 from \citet{Hamuy:1996d} and 8 constructed from well
sampled light curves, expand a large range in post-maximum decline
rates: $0.83 \leq \Delta {\rm m} _{15} (B) \leq 1.93$.  The dashed lines are
the $\pm$1$\sigma$ limits using the interpolated photometric
uncertainties as a function of time. We have applied a progressive
offset of 1 mag for plotting purposes. \label{fig_1}}

\end{figure}

In Table \ref{table_1} we present basic information of the templates
and the references of the photometric data.  All the LC templates are
plotted in Figure \ref{fig_1}, ordered by increasing values of
\mbox{$\Delta$m$_{15}(B)$} from slow decliners to fast decliners. From
Figure \ref{fig_1} we can observe the main characteristics of the
$BVRI$ light curves: a principal maximum, followed by a change in the
curvature of the light curve $\sim 10-30$ days after maximum and a
linear decay at greater than $30-40$ days. The $RI$ light curves also
have a secondary hump fainter than the first peak at $\sim 20-30$
days. A small inflection in the $V$ light curve can be seen during the
same epoch as the $I$ secondary maximum.  Different morphological
properties of the light curves, such as the time of the secondary peak
in the $RI$ filters and the time of inflection, are correlated with
the post-maximum decline rate \citep{Hamuy:1996d}. The rise times in
the $BV$ filters are generally correlated with the post-maximum
decline rate \citep{Riess:1999b}, slow risers are slow decliners, and
fast risers are fast decliners.

The dashed lines around each solid curve in Figure \ref{fig_1}
represents the approximate $\pm 1 \sigma$ statistical uncertainties in
the photometry.  These curves were obtained doing a linear
interpolation of the errors in the original data points. Typical
uncertainties in the photometry from photon statistics and errors in
the standards are \mbox{$\sim 0.02$ mag} around maximum and $\geq
0.04$ mag at late times. Note though that the systematic errors that
arise from doing photometry on a nebular (as opposed to stellar)
spectral energy distribution of late-time SNe are much larger, at the
$\sim 0.1$ mag level \citep{Suntzeff:2000}.

\section{The method}
\label{sec:2}

The principal idea of the method is a formalism which allows us to
construct a linear combination of the discrete template set, composed
by the observed templates, into a template parametrized by
\mbox{$\Delta$m$_{15}$}.  The \mbox{$\Delta$m$_{15}$} parameter can
then be included in the $\chi^{2}$ function when fitting a new
multicolor light curve, allowing us to calculate a covariance matrix
and the estimation of statistical uncertainties in the best fitting
parameters.

\subsection{Interpolation between light curve templates}
\label{sec:2.1}

We interpolate between different templates using a weighting
function. It assigns different weights to each template in the
template set, in order to obtain:

\begin{enumerate}
\item The \mbox{$\Delta$m$_{15}$} of the constructed template,  
\item The constructed template as a function of time in different filters, parametrized by \mbox{$\Delta$m$_{15}$}.
\end{enumerate}

We have engineered the weighting function to allow us to use the
information in all the templates with \mbox{$\Delta$m$_{15}$} values
{\em near} the real value, while ignoring the templates with light
curve shapes very different from the light curve we are trying to
fit. This weighting scheme will allow us to use real light curve
templates to fit the observed data, and at the same time allows us to
avoid relying on any one particular light curve template. With this
philosophy, we can drop in or remove any template without any need to
regenerate a series of interpolated (in \mbox{$\Delta$m$_{15}$})
templates. In this sense, our fitting technique tries to stay as close
to the observed template data as possible.

The shape of the weighting function should be selected according to
the template sample and their \mbox{$\Delta$m$_{15}(B)$}.  In general
we want to have $\sim 2-4$ observed templates in the interpolation to
obtain a constructed template, without introducing uncertainties in
the interpolation. For example: if we want a constructed template with
\mbox{$\Delta$m$_{15} = 1.20$ mag}, the weight assigned to the
template of \objectname{SN~1991bg}, the fastest declining LC in the
template set, should be very small or 0 because their LC shapes are
very different.

\begin{figure}
\plotone{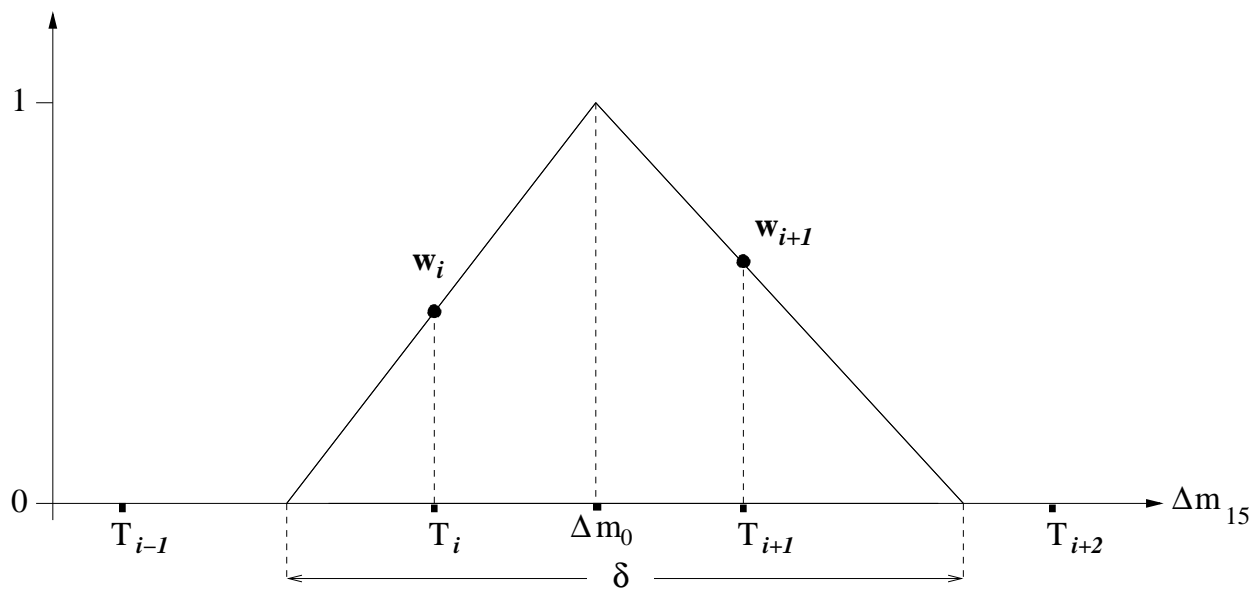}
\caption{Weighting function used for the interpolation between the
observed discrete templates in the template set. The value of the
weight assigned to each template, $w_{i}$, is equal to the value of
the triangle function in the \mbox{$\Delta$m$_{15}(B)$} of the
template $i$.  The weights are normalized so that they sum to
1. When a new light curve fitting is performed, the center of the
triangle function, $\Delta {\rm m}_{0}$, moves freely in the
\mbox{$\Delta$m$_{15}$} axis until the $\chi^{2}$ of the fit is
minimized. The final value of the \mbox{$\Delta$m$_{15}$} of the
constructed template is obtained by weighting the values of the
template set.  \label{fig_2}}
\end{figure}

We choose here a triangle as the weighting function (see Figure
\ref{fig_2}). The weight assigned to each observed template,
\mbox{w$_{i}$}, is equal to the value of the triangle function at
\mbox{$\Delta {\rm m}_{15}(B)$} of the template $i$ in the template
set:

\begin{equation}
\label{eq:1}
{\rm w} _{i} = g(\Delta {\rm m}_{0}, \Delta {\rm m }_{15}(B)_{i}) 
\end{equation}

\noindent Note that the final values of the weights are normalized so
that their total sum is 1. When the best-fit template is constructed,
the center of the triangle function, $\Delta {\rm m}_{0}$, moves along
the \mbox{$\Delta$m$_{15}$} axis until the required value of the
\mbox{$\Delta$m$_{15}$} is obtained. This value is calculated directly
by weighting the values of the observed templates:

\begin{equation}
\label{eq:3}
\Delta {\rm m} _{15} = \sum_{i=1}^{n} {\rm w}_{i}\times \Delta {\rm m}_{15}(B)_{i}
\end{equation}

\noindent The best-fit template $\vec{N}$ in the filter $X$ is 
constructed by adding the weighted observed templates $\vec{T}$ in the
template set:

\begin{equation}
\label{eq:2}
\vec{{\rm N}}^{X}(\Delta{{\rm m}}_{15}) = \sum_{i=1}^{n} {\rm w}_{i}\times  \vec{{\rm T}}^{X}_{i} 
\end{equation}  

\noindent In this case $n=14$ (see Table \ref{table_1}) is the number
of discrete templates in the template set. The vectors represent the
time dependence of the templates in different filters, $X=BVRI$,
measured in the rest frame with respect to the time of $B$ maximum.

\begin{figure}
\plotone{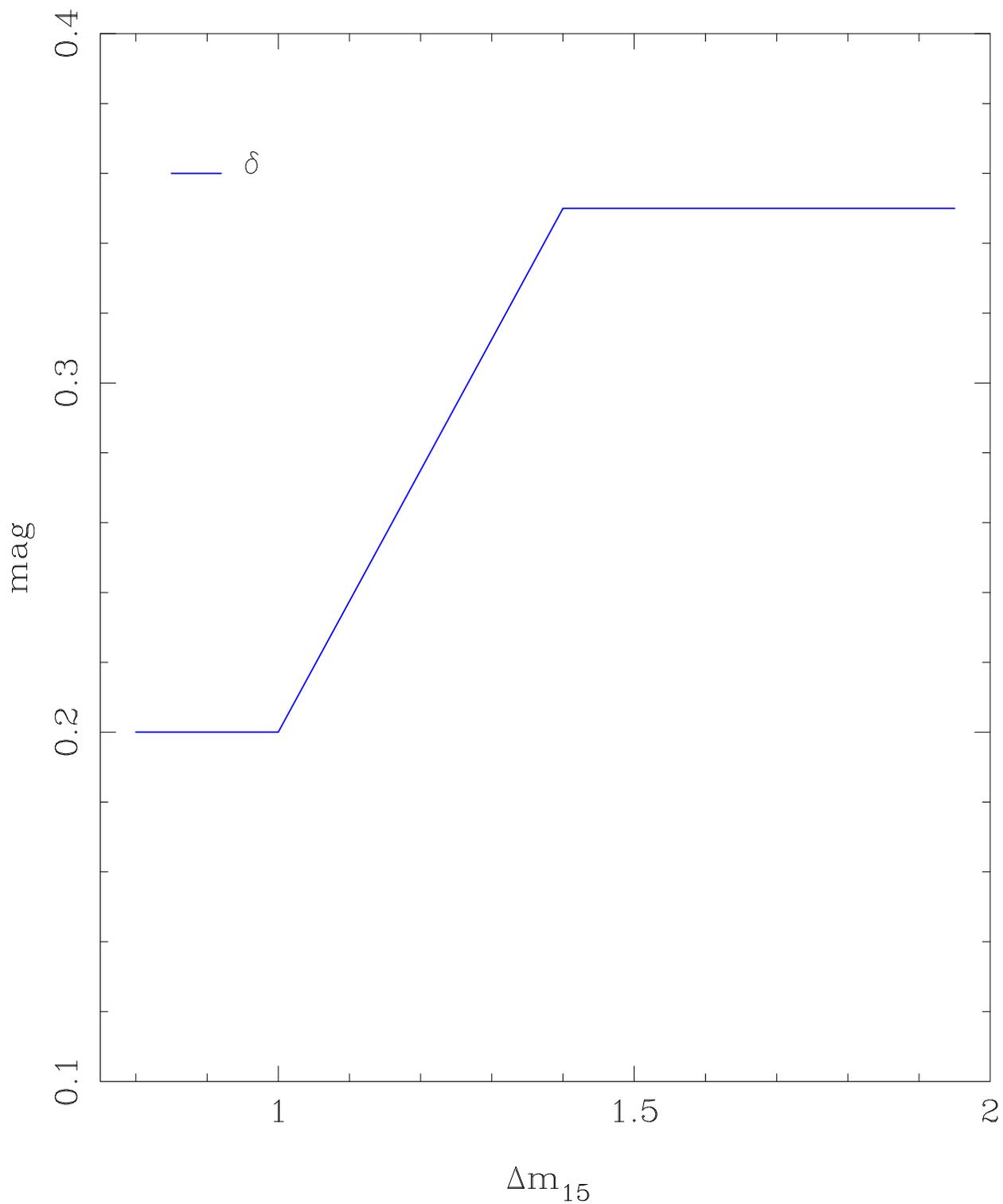}
\caption{Width of the triangle function used, $\delta$, as a function
of \mbox{$\Delta$m$_{15}$}.  The values are chosen according to the
sampling of the \mbox{$\Delta$m$_{15}(B)$} parameters of the templates
in the template sample. Since we have only 4 templates with $\Delta
{\rm m}_{15} > 1.30$ mag, we use a wider triangle function for the
fast declining templates. \label{fig_3}}
\end{figure}

In Figure \ref{fig_3} we show the value of $\delta$, the width of the
triangle function as a function of \mbox{$\Delta$m$_{15}$}. The effect
of this weighting scheme is that the discrete templates outside of the
triangle function are given 0 weight. The larger values of $\delta$
for \mbox{$\Delta$m$_{15} > 1.40$ mag} were selected to overcome the
poor sampling and small number of fast declining SNe in the template
set (see Table \ref{table_1}). 

\begin{figure}
\includegraphics[angle=0,scale=0.8]{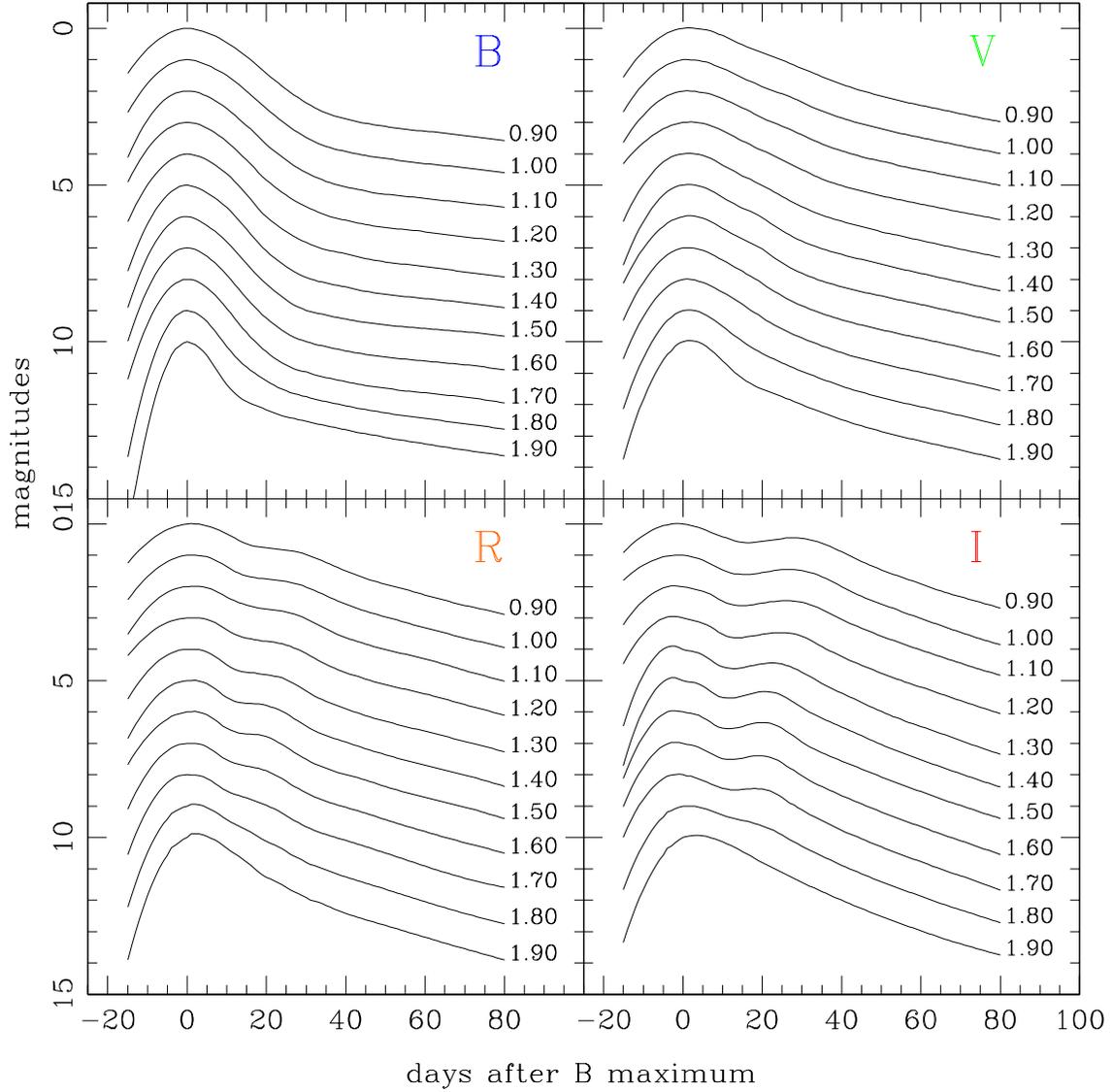}
\caption{Constructed templates with a range of values of the
\mbox{$\Delta$m$_{15}$} parameter. These set of templates result from
the interpolation of the discrete templates from the template set,
using the triangle function.  We have applied a progressive offset 
of 1 mag for plotting purposes. \label{fig_4}}
\end{figure}

In Figure \ref{fig_4} we present constructed templates with different
values of \mbox{$\Delta$m$_{15}$} as an example of the interpolation
scheme.  We have extrapolated the initial templates to earlier times
($-15 \leq {\rm t-t}_{0}(B){\rm [days]} \leq -5$) assuming a quadratic
dependence with time.  The correlation between
\mbox{$\Delta$m$_{15}$}, the post-maximum decline rate, and the rise
time is very clear in the $BV$ templates: slow risers are slow
decliners, and fast risers are fast decliners \citep{Riess:1999b}. In
the $RI$ templates this correlation is not clearly present, but it is
observed that the secondary maximum occurs later for the slow
declining LCs.

The final range of fitted \mbox{$\Delta$m$_{15}$} values is restricted
to the range in the input template set. For the template set presented
in our work: \mbox{$0.83 \leq$} \mbox{$\Delta$m$_{15}{\rm \, [mag]}
\leq 1.93 $} (Table \ref{table_1}). 

\subsection{$\chi^{2}$ fitting}
\label{sec:2.2}

In the original method \citep{Phillips:1993, Hamuy:1996a, Hamuy:1996b,
Phillips:1999} and the modified version of \citet{Germany:2004},
\mbox{$\Delta$m$_{15}$} is not fitted directly in one step. When a new
LC is fitted, the total reduced $\chi^{2}$ is calculated for each
template summed across all colors. The \mbox{$\Delta$m$_{15}$} of the
SN is derived with a second-order polynomial fit to the curve
\mbox{$\Delta$m$_{15}$} versus $\chi^{2}_{\nu}$, selecting the value
where $\chi^{2}_{\nu}$ reaches a minimum as the best fitting
parameter. This technique is justified by the fact that $\chi^{2}$
function changes quadratically with the parameters near the minimum
\citep{Bevington, Press:1988}.  This approach has two problems:

\begin{itemize}

\item the time of $B$ maximum, $BVRI$ magnitudes at maximum and
\mbox{$\Delta$m$_{15}$} are correlated parameters. This correlation is
not considered in the quadratic fit to $\chi^{2}$.

\item the statistical uncertainties in the best fitting value of
\mbox{$\Delta$m$_{15}$} are difficult to estimate in a consistent way.

\end{itemize}

The linear combination of templates weighted as described above,
allows us to include \mbox{$\Delta$m$_{15}$} directly in 
the $\chi^{2}$ function.  In this way, the most general version of the
empirical model for a new LC in the observed filter $Y$ is:

\begin{eqnarray}
\label{eq:4}
\vec{\mathcal{M}}^{Y} = \vec{\rm N}^{X}(\Delta {\rm m}_{15}) + M_{max}^{X}(\Delta {\rm m}_{15}) + \mu_{0} 
\\+ \vec{K}_{XY}(\Delta {\rm m}_{15},E(B-V)) + \mathcal{R}_{X}\,E(B-V)_{host} + \mathcal{R}_{Y}\,E(B-V)_{Gal}  \nonumber
\end{eqnarray}

\noindent where: $M_{max}^{X}(\Delta {\rm m}_{15})$ are the relations
between the absolute magnitude at maximum in the rest frame filter
$X=BVRI$ and $\Delta {\rm m}_{15}$ (see Section \ref{sec:3} for the
derivation of this relations); $\mu_{0}$ is the reddening corrected
distance modulus of the SN; $\vec{K}_{XY}(\Delta {\rm m}_{15},E(B-V))$
are the cross-filter $K$-corrections \citep{Kim:1996, Schmidt:1998,
Nugent:2002, Germany:2004}, which depends on the colors of the SN and
total extinction; $\mathcal{R}_{X}$ and $\mathcal{R}_{Y}$ are the
ratio of total to selective extinction in the host galaxy and in the
Milky Way, respectively; $E(B-V)_{host}$ and $E(B-V)_{Gal}$ are the
color excesses in the host galaxy and in the Milky Way
\citep{Schlegel:1998}. 

When fitting a new SN LC we use this empirical model to construct a
multidimensional $\chi^{2}$ function:

\begin{equation}
\label{eq:5}
\chi^{2}({\rm t}_{0}, \Delta {\rm m}_{15}, E(B-V)_{host}, \mu_{0}) = [\vec{\mathcal{M}}^{Y} - \vec{m}^{Y}]^{T} ({\bf C}^{Y})^{-1} [\vec{\mathcal{M}}^{Y} - \vec{m}^{Y}]
\end{equation} 

\noindent where $(\vec{\mathcal{M}}^{Y} -\vec{{m}}^{Y})$ are the
residuals of the fit; $\vec{m}^{Y}$ are the observed magnitudes in the
$Y$ filter (light curve data); ${\rm t}_{rest frame} = ({\rm
t}_{obs}-{\rm t}_{0}(B))/(1+z)$, with $z$ the cosmological redshift,
is the rest frame time which parametrizes the vectors; ${\bf C}^{Y}$ is the
correlation matrix \citep{Riess:1996}.  This symmetric matrix contains
the variances (squared statistical uncertainties) of the data for a given
date (diagonal terms) and the correlation between data of
different days in the same filter (non-diagonal terms). We use a
diagonal matrix for ${\bf C}^{Y}$, setting the non-diagonal elements to
zero, where each element takes into account the uncertainties in the
data, the model, and the $K$-corrections: $\sigma_{data}^{2} +
\sigma_{model}^{2} + \sigma_{K-corr}^{2}$.  The uncertainties in the
model, $\sigma_{model}$, are calculated by weighting the uncertainties 
in the observed templates with the same weights calculated from 
Equation \ref{eq:1}.

The $\chi^{2}$ function of Equation \ref{eq:5} is minimized when a new
SN light curve is fitted. This multidimensional (i.e. multicolor)
minimization gives the best values for the parameters of the empirical
model of Equation \ref{eq:4}. The fitting parameters of the model are:
time of $B$ maximum, \mbox{$\Delta$m$_{15}$}, color excess of the host
galaxy, and the reddening corrected distance modulus.

Several works in the literature suggest that the average values of
selective to total extinction of host galaxies are smaller than the
Galactic values \citep{Riess:1996, Tripp:1999, Phillips:1999,
Wang:2003, Altavilla:2004, Riendl:2005}.  Also $\mathcal{R}$ changes
in time because of the fast evolution of the SED of Type Ia SNe
\citep{Leib:1988, Phillips:1999, Nugent:2002, Jha:2006b}, producing a
dependence of $\Delta {\rm m}_{15}$ with the total extinction towards
the SN.  We have restricted in this work to constant values of
$\mathcal{R}$, equal to the Galactic reddening law
\citep{Cardelli:1989}: $\mathcal{R}_{B}$=4.20, $\mathcal{R}_{V}$=3.10,
$\mathcal{R}_{R}$=2.54, $\mathcal{R}_{I}$=1.85. In a future paper we
will explore fitting the extinction law as part of the generalized
$\chi^{2}$ fit.

To summarize, Equations \ref{eq:2} through \ref{eq:5} allow us to
introduce a linear combination of templates directly into the
$\chi^{2}$ fitting procedure. We do not have to output an intermediate
table of interpolated templates (such as those shown in
Figure \ref{fig_4}). The mathematical framework allows us to change the
weighting, the template set, and even the parameters to be fit in an
elegant way.

\subsection{Uncertainties in the fitting parameters}
\label{sec:2.3}

We calculate the uncertainties in the parameters of the best fitted LC
model by constructing the covariance matrix. If we have
$\chi^{2}(a_{i})$, with $a_{i}$ as the fitting parameters, the elements
of the covariance matrix ${\bf \alpha}$ are \citep{Bevington,
Press:1988}:

\begin{equation}
\label{eq:6}
\alpha_{l,k} = \frac{1}{2} \frac{\partial^{2} \chi^{2}}{\partial a_{l} \partial a_{k}} 
\end{equation}
                                                                                            
\noindent The partial derivatives are evaluated in the best fit
parameters $a_{i}^{best}$. The inverse of the covariance matrix, the
error matrix ${\bf \epsilon} = {\bf \alpha^{-1} }$, has the variances
(diagonal terms) and covariances (non-diagonal terms) in the best
fitting parameters. The final $1\sigma$ uncertainties in the fitting
parameters are estimated using the following approximate formula
\citep{Press:1988}:

\begin{equation}
\label{eq:7}
\sigma _{a_{i} ^{best}} = \sqrt{ \chi^{2}_{\nu} \cdot \epsilon_{i,i}}
\end{equation}

\noindent where: $\chi^{2}_{\nu}$ is the value of the minimum
$\chi^{2}$ divided by the number of degree of freedom of the fit $\nu
= N - m$, with $N$ the total number of data points fitted and $m$ the
number of fitting parameters ($m=4$ in current implementation).

\section{Calibration of the relations between $M_{max}$ and $\Delta {\rm m}_{15}$} 
\label{sec:3}

The measurement of distances to Type Ia SNe using the
\mbox{$\Delta$m$_{15}$} method is based on the observed correlations
between the absolutes magnitudes at maximum light \mbox{($M_{max}$)}
and the rate of evolution away from maximum light as parametrized
originally with \mbox{$\Delta$m$_{15}$} \citep{Phillips:1993,
Hamuy:1996a, Phillips:1999, Germany:2004} established with a low
redshift sample of SNe in different rest frame filters (second term in
the right hand side of Equation \ref{eq:4}).  Different versions of
this relations have been proposed in the literature: linear relations
\citep{Phillips:1993,Hamuy:1996a}, quadratic polynomials
\citep{Phillips:1999, Germany:2004}, both restricted to
\mbox{$\Delta$m$_{15} \leq 1.70$ mag}, and an exponential that
includes the fast declining light curves
\citep{Garnavich:2004}. Finally, a relation that includes both light
curve shape and color (as a second parameter) has been published by
\citet{Parodi:2000}.

We used a modified version of the analytical model of Equation
\ref{eq:4} to fit the multicolor light curves of a large sample of
Type Ia SNe at $z < 0.12$ (see Table \ref{table_2}) in order to study
the relations between $M_{max}$ and \mbox{$\Delta$m$_{15}$} in
different filters.  The parameters of this multicolor fits are: rest
frame apparent $BVRI$ magnitudes at maximum corrected by Galactic and
host reddening, \mbox{$E(B-V)_{host}$}, \mbox{$\Delta$m$_{15}$}, and
time of maximum in $B$. This methodology is similar to the one applied
by \citet{Germany:2004} (section 4.4.4) in their modified $\Delta {\rm
m}_{15}$ method. The main difference is that they used the unreddened
sample of Type Ia SNe defined by \citet{Phillips:1999}, whereas we use
a larger sample of SNe in a self consistent way analyzed with the
analytical technique introduced in Section \ref{sec:2} to calculate
$\Delta {\rm m}_{15}$ and $E(B-V)_{host}$ simultaneously from the
multicolor light curves.

Since the magnitudes at maximum light and the host galaxy reddening
are highly correlated parameters in the analytical model, priors are
needed to soften this degeneracy. We applied them directly in the
$\chi^{2}$ function as in \citet{Jha:2006b}. First, negative values of
$E(B-V)_{host}$ were handled using a Bayesian filter
\citep{Riess:1996}, assuming a one sided Gaussian a priori
distribution of $A_{B}$ with maximum at $A_{B}=0$ and
$\sigma(A_{B})=0.3$ mag \citep{Phillips:1999}. Secondly,
\citet{Jha:2006b} studied the intrinsic colors of a large sample of
Type Ia SNe and found that their colors at 35 days after $B$ max are
very homogeneous and well described by a Gaussian distribution with
$(B-V)_{35}=1.055$ mag and $\sigma=0.055$ mag. We applied this as a
prior in the reddening corrected $B-V$ colors, which is very similar
to the ``Lira law'' \citep{Lira:1995}. Because it is outside the scope
of this paper, we leave as an open question how the host galaxy
reddening obtained from the fits depends on the priors.

We assumed a concordance cosmology with
$(h,\Omega_{M},\Omega_{\Lambda})=(0.72,0.3,0.7)$ to transform the rest
frame apparent magnitudes to absolute magnitudes for SNe in the Hubble
flow at redshifts $z>0.01$ (Cal\'an/Tololo, CfAI, Krisciunas et~al. and 
CfAII subsamples in Table \ref{table_2}).  For the nearby
sample we used the distance moduli to the host galaxies obtained with
the Surface Brightness Fluctuation method \citep{Ajhar:2001}, matching
the derived distances in $H_{0}=75 [{\rm km\, s^{-1}\, Mpc^{-1}}]$
scale to $H_{0}=72 [{\rm km\, s^{-1}\, Mpc^{-1}}]$.  The final error
budget in the absolute magnitudes includes: statistical uncertainties
in the reddening corrected apparent magnitudes from the covariance
matrix and a Hubble flow ``noise'' of \mbox{600 [km s$^{-1}$]} to
crudely model the effects of peculiar velocities \citep{Marzke:1995}.
For the nearby sample we only included the error in the distance
modulus of the host galaxies.

\begin{figure}
\includegraphics[scale=0.85]{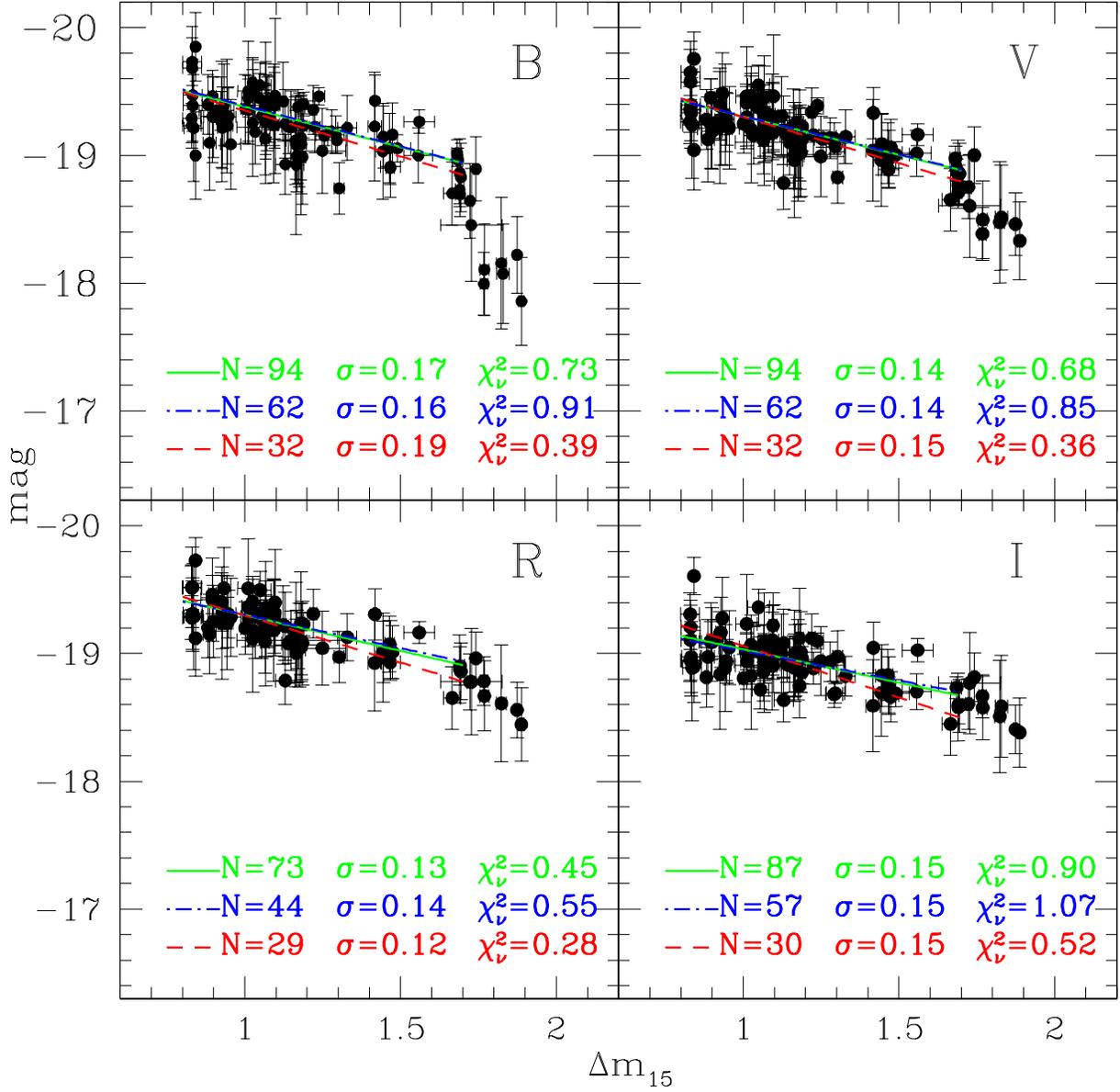}
\caption{Relations between the absolute magnitudes at maximum and
$\Delta {\rm m}_{15}$ in $BVRI$ filters. The lines are linear fits
performed in the range $0.80 < \Delta {\rm m}_{15} {\rm [mag]} < 1.70$
to the complete sample (solid) of Table \ref{table_2}, unreddened
sample with \mbox{$E(B-V)_{host} \leq 0.06$} mag (dot-dashed), and the
reddened sample with $E(B-V)_{host} > 0.06$ mag
(dashed). \label{fig_5}}

\end{figure}

In Figure \ref{fig_5} we plot the results of the fits to the complete
sample of Type Ia SNe listed Table \ref{table_2}.  For the SNe that
are in common between Krisciunas et al. and CfAII sample (4 SNe), we
have selected the results of the fits with smaller $\chi^{2}_{\nu}$
for the complete sample. The lines are linear fits to the relations
between $\Delta m_{15}$ and $M_{max}$ in the range $0.80 \leq \Delta
m_{15}\leq 1.70$. We use $E(B-V)_{host}=0.06$ mag as a cut in host
reddening for comparison with other works in the literature
(e.g. \citet{Phillips:1999}). In Table \ref{table_3} we present the
results of the various linear fits in different filters to the
complete sample.

We restrict the linear fits to $\Delta m_{15}= 1.70$ mag mainly
because this cut has been widely used in other works of $\Delta
m_{15}$ method (e.g. \citet{Phillips:1999}). \citet{Garnavich:2004}
included subluminous (fast-declining) SNe to obtain distances, but the
dispersion in the Hubble diagram increases and the sample of
subluminous SNe is still small and poorly studied compared with normal
SNe Ia.

We note that quadratic fits to the data, as the ones applied by
\citet{Phillips:1999}, did not improve the reduced $\chi^{2}$ in this
range of decline rates. Also, we did not take out SNe identified as
outliers in previous papers because we cannot determine, in an
unbiased way, if they are outliers in our analysis.

There are several important points to note about Figure
\ref{fig_5}. The dispersion around the linear fits does not correlate
with wavelength, although the dispersion in $B$ is slightly bigger than
in $VRI$ filters. This is probably due to a few SNe that deviate
substantially from the whole sample, as it is evident in the
Figure. This could be caused by a different reddening law in the host
galaxies of this SNe, but varying the reddening law was outside the
scope of the analysis presented here.

The zero-points of the linear relations are consistent, within the
uncertainties, when we split the sample by host galaxy reddening. The
slopes are marginally inconsistent, mainly in $R$ and $I$ filters, and
steeper in the high-reddening sample. We think this is produced by the
smaller number of fast declining SNe \mbox{($ 1.50 < \Delta {\rm
m}_{15} < 1.70$)} in the sample. The relations obtained here are
consistent with previous works in the literature that include
corrections for host galaxy reddening \citep{Phillips:1999,
Germany:2004}.

In almost all the cases $\chi^{2}_{\nu} \lesssim 1.0$ in the linear
fits, which could be produced by an overestimation of uncertainties in
the absolute magnitudes. $\chi^{2}_{\nu}$ depends strongly in the
adopted ``noise'' in the Hubble flow due to peculiar velocities which
has values between 250-600 \mbox{[km s$^{-1}$]} in the literature
\citep{Baker:2000,Landy:2002,Zehavi:2002, Hawkins:2003}. If we had 
chosen 400 \mbox{[km s$^{-1}$]} for instance, the values of $\chi^{2}_{\nu}$ 
would be $\sim 1.5$ times higher. 

The dispersion around the linear fits seems to be larger at the lower
end for the slowly declining SNe. This could be indication of a higher
intrinsic dispersion in the slow decliners as noted by other authors.
However, it is possible that the $\Delta {\rm m}_{15}$ values of some of these
slow SNe are smaller than the minimum $\Delta {\rm m}_{15}$ of SNe in
the template set. More well-sampled light curves of slow decliners are
needed to extend and improve the template set in order to study the
apparent larger natural dispersion.

In Table \ref{table_4} we present the linear relations obtained for
different subsamples in Table \ref{table_2}. The zero-points of the
linear relations are consistent, within the 1$\sigma$ uncertainties,
in $RI$ filters. This is not the case in $BV$ filters where there is a
clear difference (up to $\sim$0.17 mag in $B$) between CfAI,
Krisciunas et. al and CfAII, which have consistent values, and
Cal\'an/Tololo which is brighter. The nearby sample is consistent with
Cal\'an/Tololo. This zero-point difference roughly represents the magnitude
difference at $\Delta {\rm m}_{15} = 1.1$ mag.

This same difference persists if we take the weighted average
difference in the range $1.0 \leq \Delta {\rm m}_{15}\leq 1.2$. This
difference is $\sim 0.10\pm 0.03$ mag in $BV$ filters, with
Cal\'an/Tololo magnitudes again being brighter.

It is well known, but poorly studied, that photometry of SNe is
sensitive to the throughput system of the telescope/filter/detector
due the non-stellar nature of SN spectra
\citep{Suntzeff:2000}. \citet{Riess:1999a} showed (their Table 1) that
there are systematic differences in the observed photometry of SN1994D
between the FLWO Observatory (CfAI and CfAII surveys) and CTIO data
(Cal\'an/Tololo survey).  These differences are larger in the $B$
filter but always \mbox{$< 0.04$ mag}, which are significantly less
than the differences seen in absolute magnitudes at $\Delta {\rm
m}_{15} = 1.1$.  However, the comparison of the absolute magnitudes in
different data sets is not as simple because the absolute magnitudes
at maximum are obtained from the fitting, which involves K-corrections
and extinction corrections.

The systematic differences could arise from the poorly measured
telescope/filter/detector transmission functions which define the
natural photometric system. With accurate transmission functions and
accurate atlases of spectrophotometry of SNe, we could calculate the
S-corrections \citep{Stritzinger:2002} needed to bring the photometry
of non-stellar SEDs onto a photometric standard system.

Another effect that could be introducing zeropoint shifts between
subsamples is the existence of a Hubble bubble \citep{Zehavi:1998}, a
systematically lower expansion rate of $\sim 7\%$ measured for SNe Ia
at $cz\geq 7400$ \mbox{[km s$^{-1}$]} \citep{Jha:2006b}. If we assume
that the Hubble bubble alone is producing the systematic differences,
the difference in zeropoints between Cal\'an/Tololo and CfAI-II
subsamples would be 0.06-0.09 mag. This effect does not fully explain
the differences that we find.

The slopes of the linear fits in different subsamples, listed in Table
\ref{table_4}, correct the SN peak magnitudes to a standard candle
value. The slopes are consistent within the errors for Cal\'an/Tololo,
CfAI and Krisciunas et al. The shallower slopes of CfAII and nearby
samples are produced by the SNe at high values of $\Delta
\rm{m_{15}}$. Removing the SNe with $\Delta \rm{m_{15}} > 1.5$ mag
brings the slope of the CfAII sample into agreement with the fits to
the other samples.

The dispersion around the linear fits are different between
subsamples. For instance, the average dispersions in Cal\'an/Tololo
sample are smaller than the other subsamples (see Table
\ref{table_4}. While the dispersion of Cal\'an/Tololo for different
filters is between $\sigma=0.11-0.14$ mag, CfAI and CfAII are in the
range $\sigma=0.11-0.19$ and $\sigma=0.15-0.17$, respectively.

Part of the difference in dispersions could be due to the fact that
the subsamples have different redshift coverage. We can compare the
intrinsic dispersions subtracting off the effect of the random
peculiar velocity of galaxies at the mean redshift of each subsample
from the dispersions in Table \ref{table_4}, this is $\sigma_{int}^{2}
= \sigma^{2} - \sigma_{v}^{2}$. The mean redshift of Cal\'an/Tololo
sample is $\bar{z}\simeq 0.05$ while for CfAI-II is $\bar{z}\simeq
0.02-0.03$. If we assume a random peculiar velocity of
$\sigma_{v}$=400 \mbox{[km s$^{-1}$]} the intrinsic dispersions in the
$B$ filter of Cal\'an/Tololo, CfAI and CfAII are consistent within
0.03 mag, with $\sigma_{int}=0.09-0.12$ mag. This result is strongly
dependent on the value of the peculiar velocity used.

The different dispersions and the different absolute magnitudes at
$\Delta \rm{m_{15}} = 1.1$ among the data subsamples point out the
urgency in producing a new set of uniform SN light curves, taken
preferably on a single telescope with a well calibrated
telescope/filter/detector system. However, a more detailed analysis is
needed to properly model the effects of a Hubble bubble in the
zeropoints and the peculiar velocity field of galaxies, coupled with
the real redshift distributions of subsamples, this could explain in
part the differences observed.

\section{Hubble diagram of low-$z$ Type Ia SNe}
\label{sec:4}

\begin{figure}
\plotone{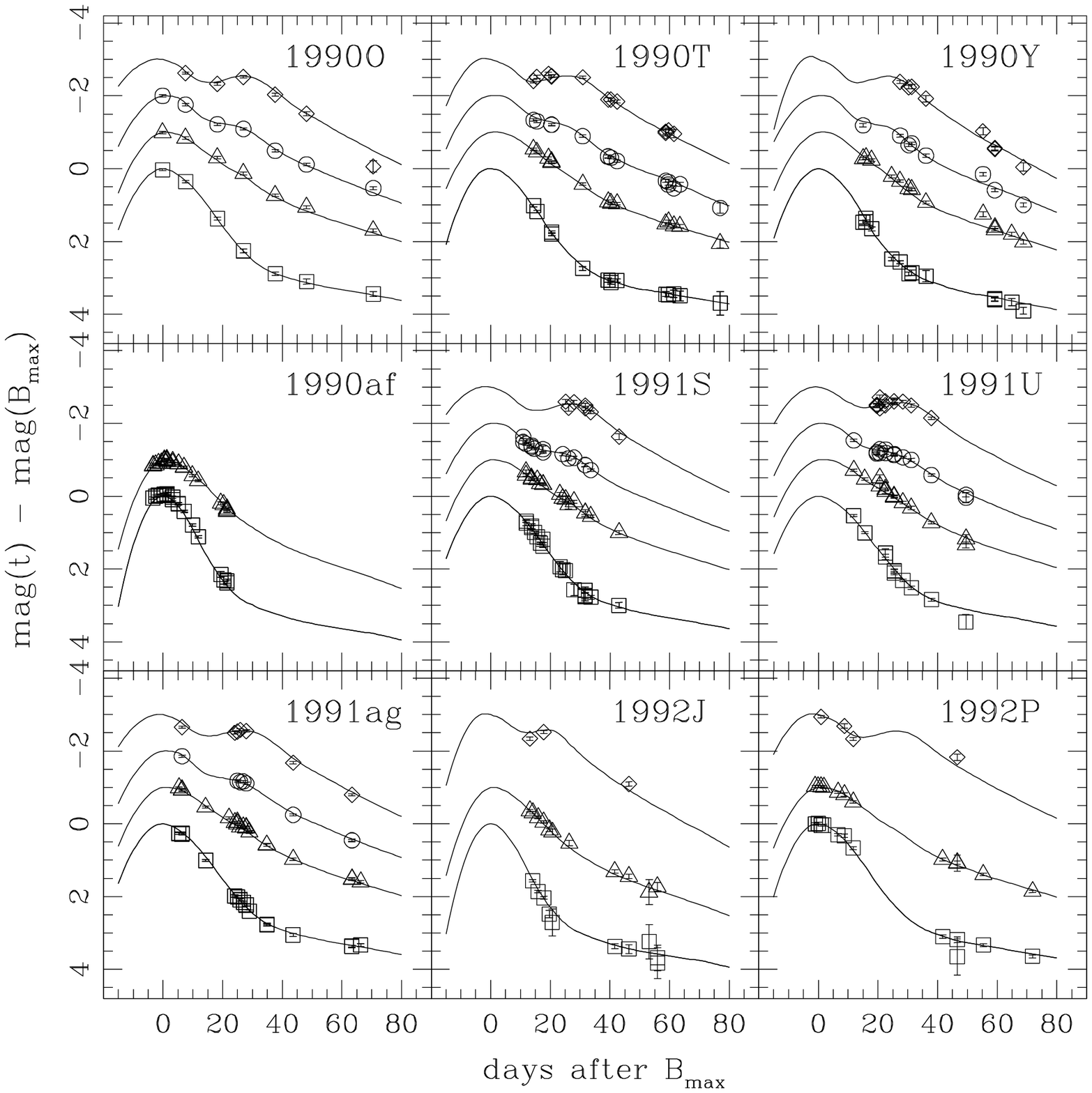}
\caption{Results of the fits to the multicolor light curves. The
different filters (symbols) are: $B$ (squares), $V$ (triangles), $R$
(circles) and $I$ (rhombuses).  To avoid overlap between different
filters we applied shifts to the relative magnitudes: $\Delta B$,
$\Delta V$-1, $\Delta R$-2, $\Delta I$-3. \label{fig_7}}
\end{figure}

\begin{figure}
\plotone{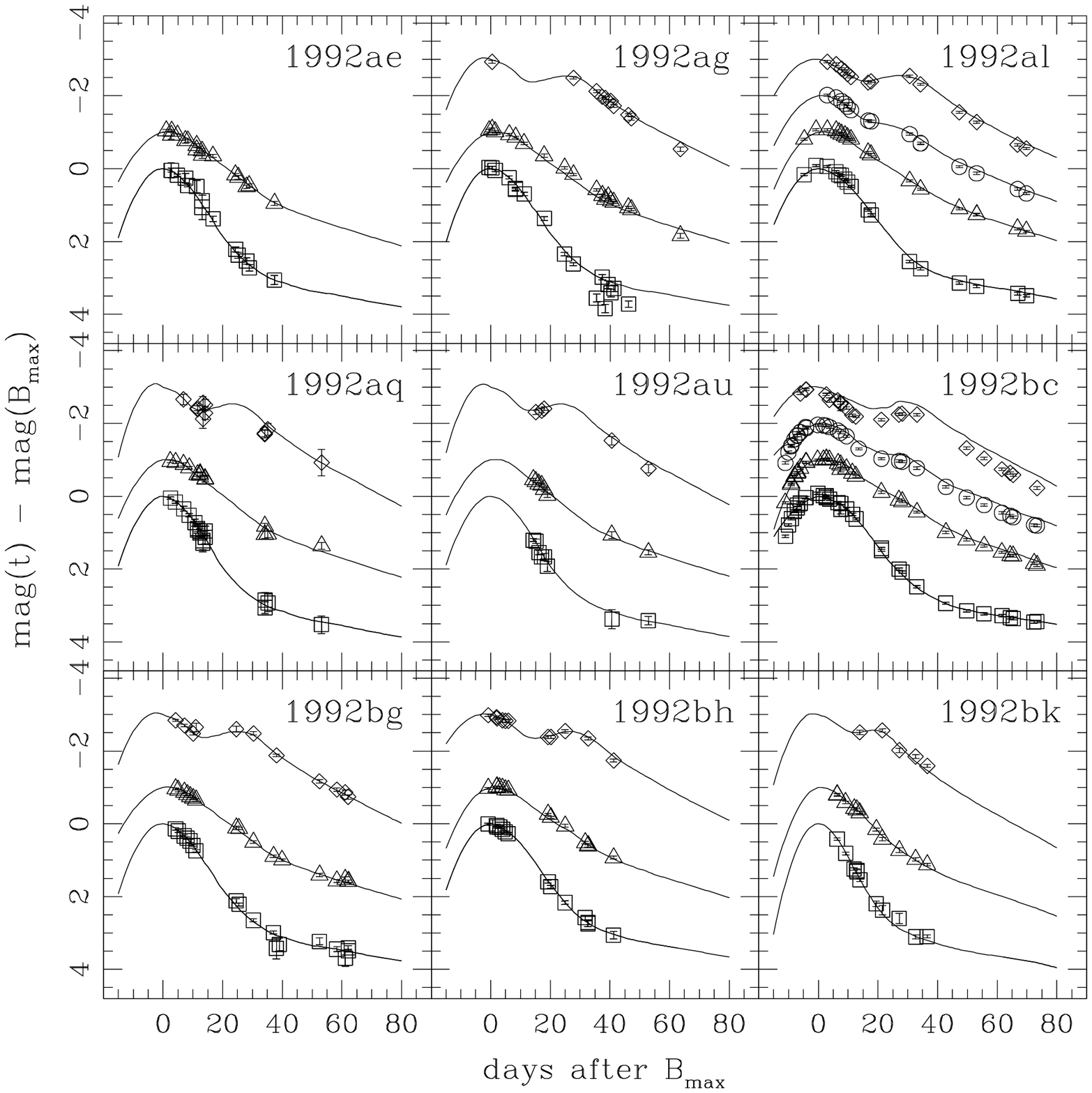}
\caption{Results of the fits to the multicolor light curves. The
different filters (symbols) are: $B$ (squares), $V$ (triangles), $R$
(circles) and $I$ (rhombuses).  To avoid overlap between different
filters we applied shifts to the relative magnitudes: $\Delta B$,
$\Delta V$-1, $\Delta R$-2, $\Delta I$-3. \label{fig_8}}
\end{figure}

\begin{figure}
\plotone{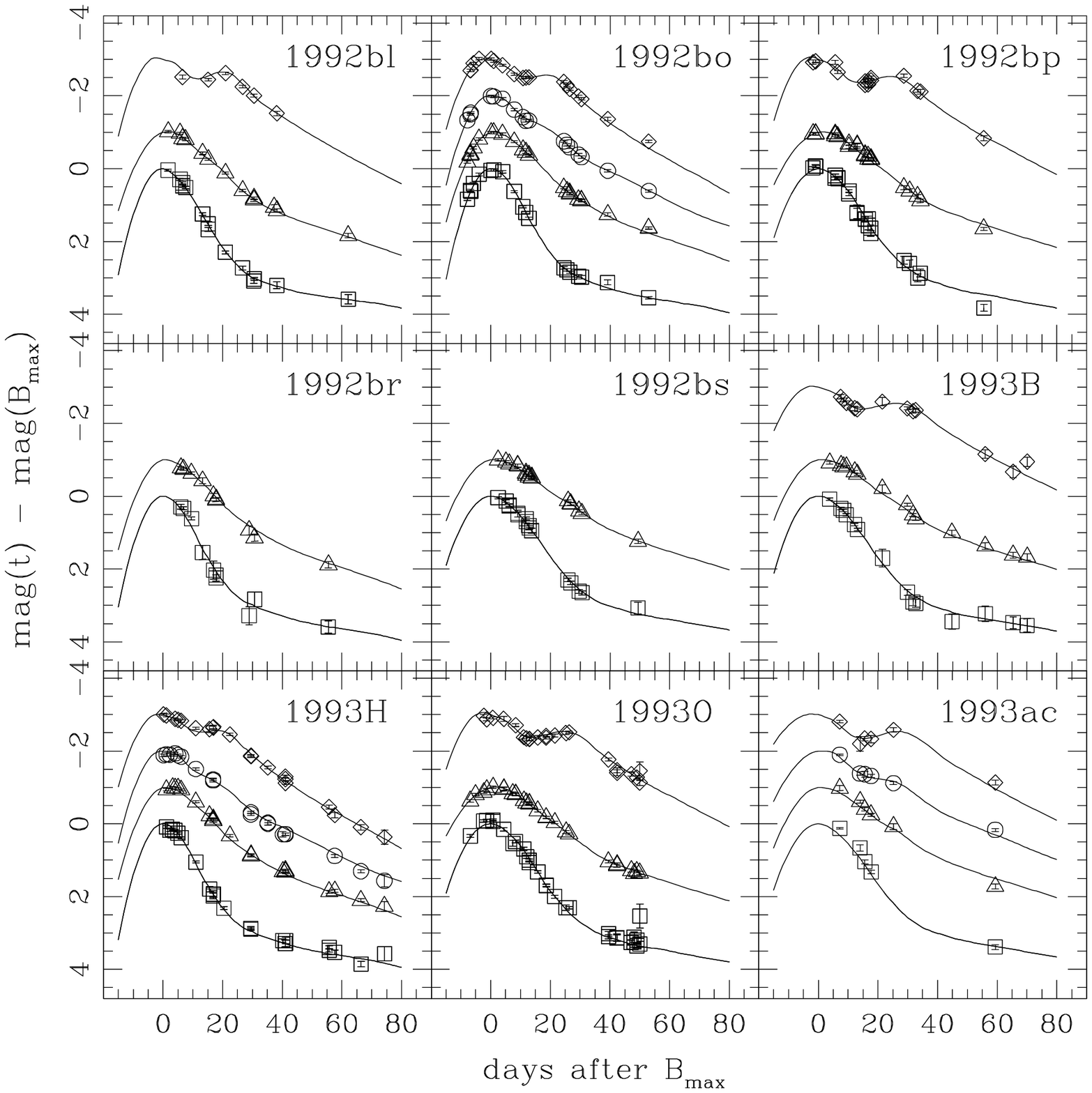}
\caption{Results of the fits to the multicolor light curves. The
different filters (symbols) are: $B$ (squares), $V$ (triangles), $R$
(circles) and $I$ (rhombuses).  To avoid overlap between different
filters we applied shifts to the relative magnitudes: $\Delta B$,
$\Delta V$-1, $\Delta R$-2, $\Delta I$-3. \label{fig_9} }
\end{figure}

\begin{figure}
\plotone{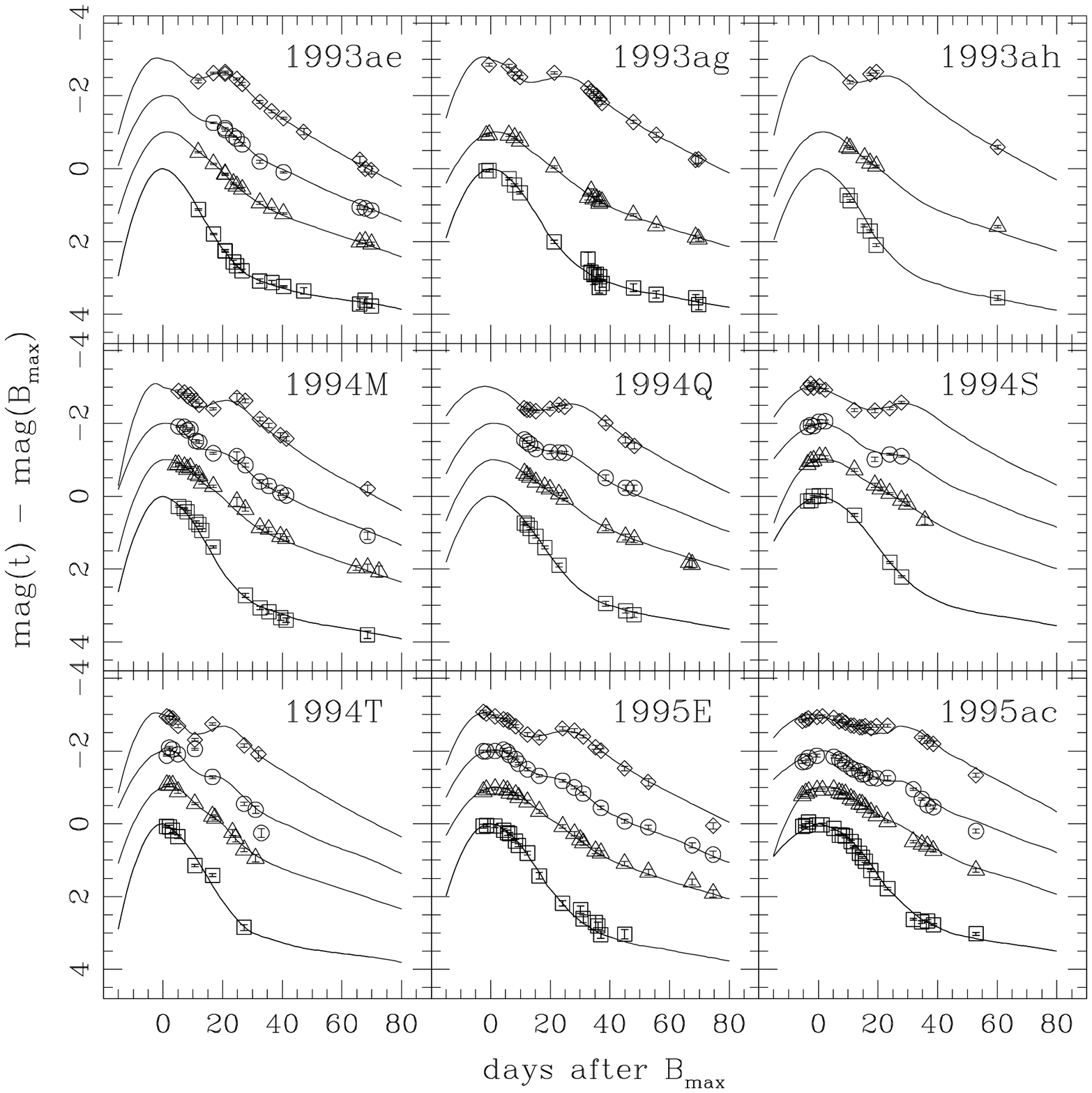}
\caption{Results of the fits to the multicolor light curves. The
different filters (symbols) are: $B$ (squares), $V$ (triangles), $R$
(circles) and $I$ (rhombuses).  To avoid overlap between different
filters we applied shifts to the relative magnitudes: $\Delta B$,
$\Delta V$-1, $\Delta R$-2, $\Delta I$-3. \label{fig_10}}
\end{figure}

\begin{figure}
\plotone{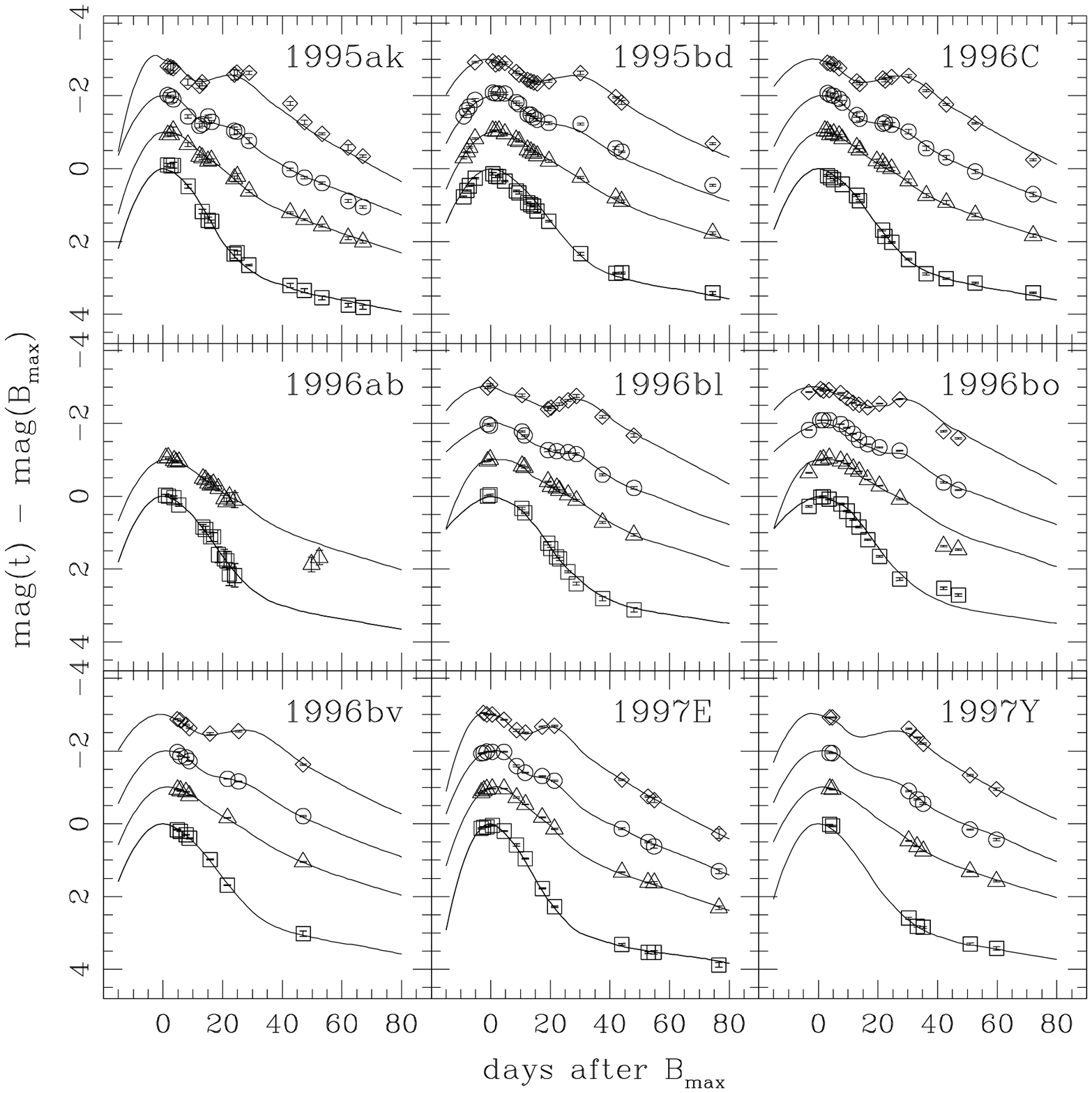}
\caption{Results of the fits to the multicolor light curves. The
different filters (symbols) are: $B$ (squares), $V$ (triangles), $R$
(circles) and $I$ (rhombuses).  To avoid overlap between different
filters we applied shifts to the relative magnitudes: $\Delta B$,
$\Delta V$-1, $\Delta R$-2, $\Delta I$-3. \label{fig_11}}
\end{figure}

\begin{figure}
\plotone{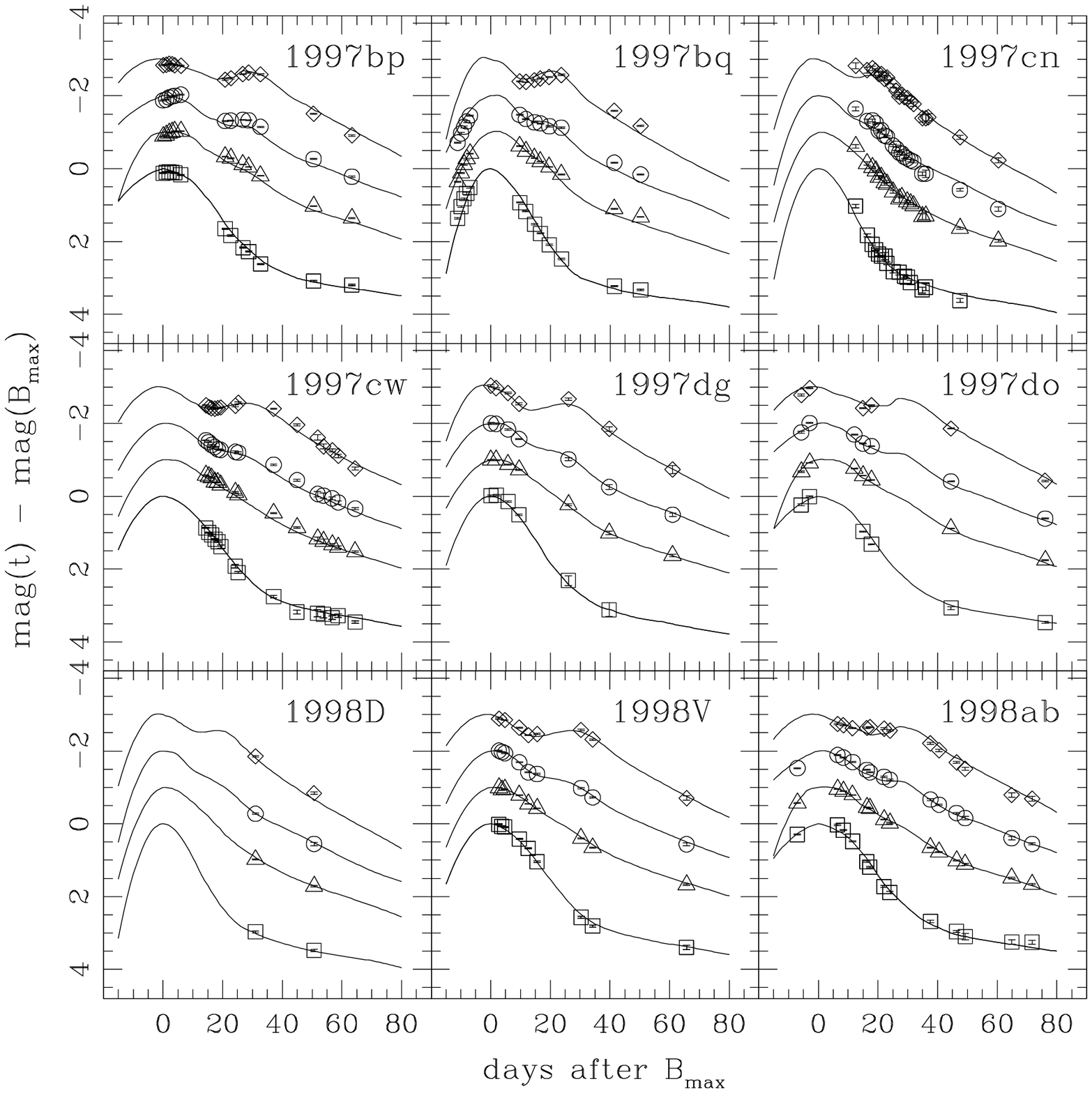}
\caption{Results of the fits to the multicolor light curves. The
different filters (symbols) are: $B$ (squares), $V$ (triangles), $R$
(circles) and $I$ (rhombuses).  To avoid overlap between different
filters we applied shifts to the relative magnitudes: $\Delta B$,
$\Delta V$-1, $\Delta R$-2, $\Delta I$-3. \label{fig_12} }
\end{figure}

\begin{figure}
\plotone{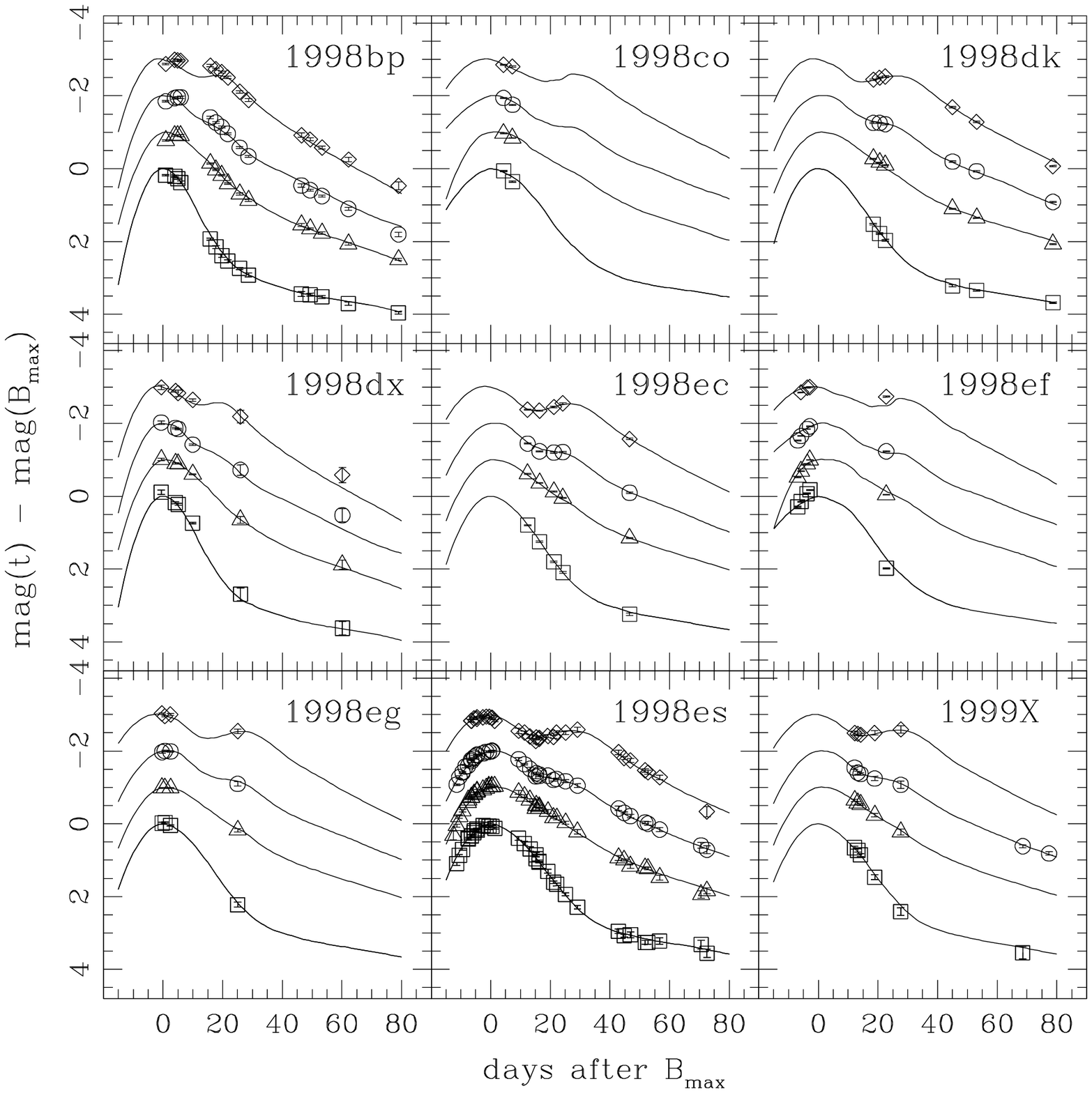}
\caption{Results of the fits to the multicolor light curves. The
different filters (symbols) are: $B$ (squares), $V$ (triangles), $R$
(circles) and $I$ (rhombuses).  To avoid overlap between different
filters we applied shifts to the relative magnitudes: $\Delta B$,
$\Delta V$-1, $\Delta R$-2, $\Delta I$-3. \label{fig_13} }
\end{figure}

\begin{figure}
\plotone{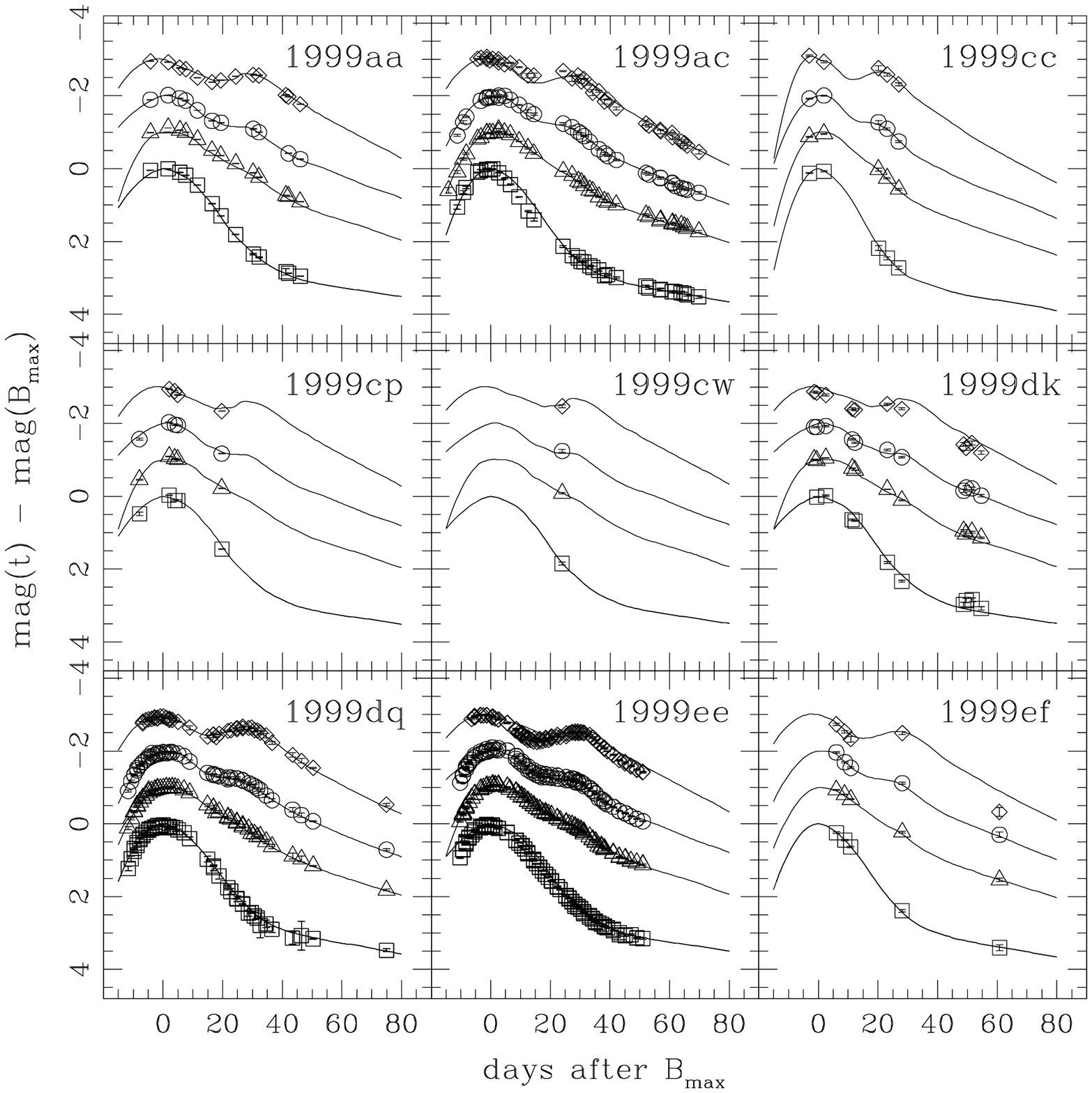}
\caption{Results of the fits to the multicolor light curves. The
different filters (symbols) are: $B$ (squares), $V$ (triangles), $R$
(circles) and $I$ (rhombuses).  To avoid overlap between different
filters we applied shifts to the relative magnitudes: $\Delta B$,
$\Delta V$-1, $\Delta R$-2, $\Delta I$-3. \label{fig_14}}
\end{figure}

\begin{figure}
\plotone{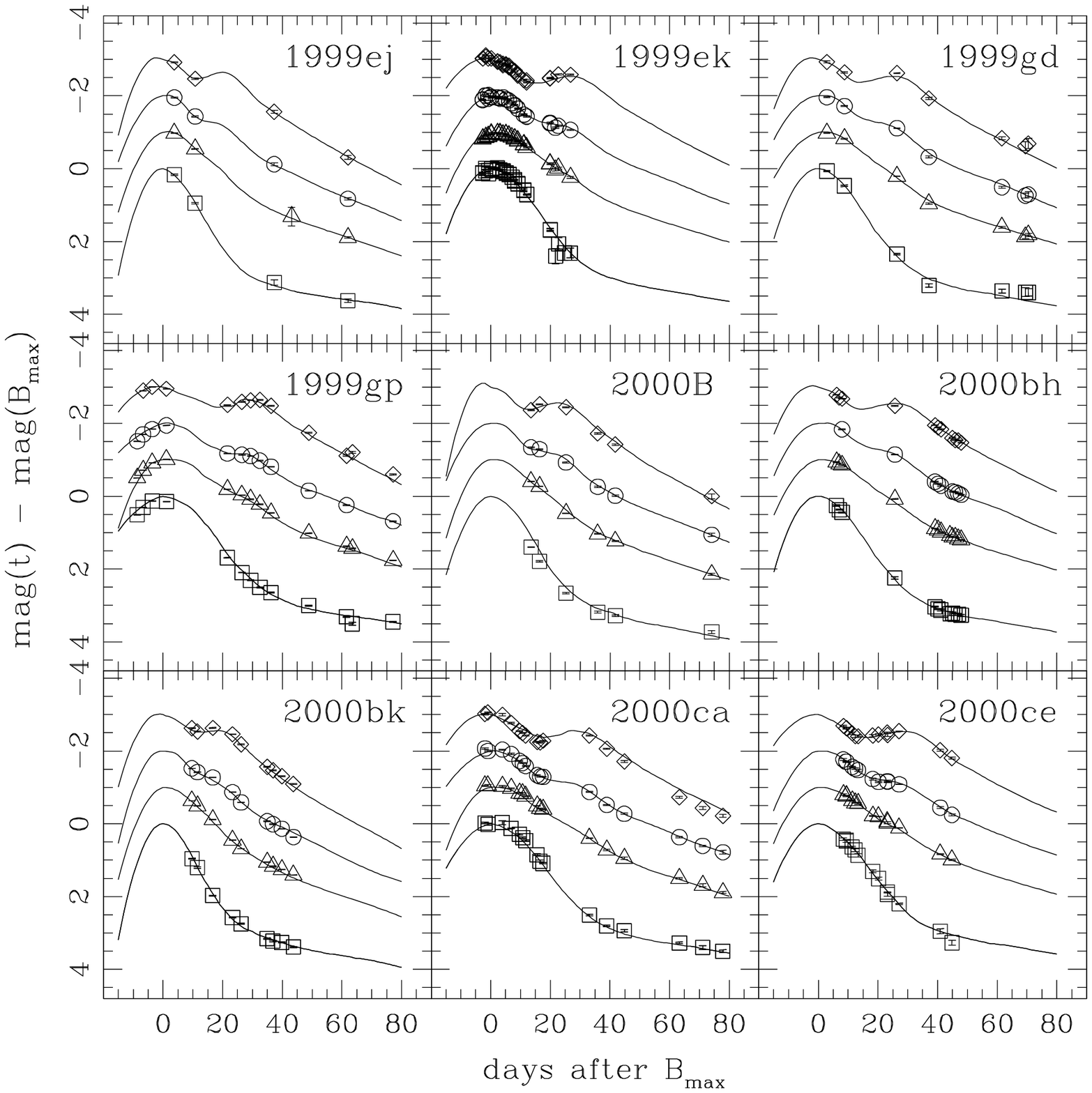}
\caption{Results of the fits to the multicolor light curves. The
different filters (symbols) are: $B$ (squares), $V$ (triangles), $R$
(circles) and $I$ (rhombuses).  To avoid overlap between different
filters we applied shifts to the relative magnitudes: $\Delta B$,
$\Delta V$-1, $\Delta R$-2, $\Delta I$-3. \label{fig_15}}
\end{figure}

\begin{figure}
\plotone{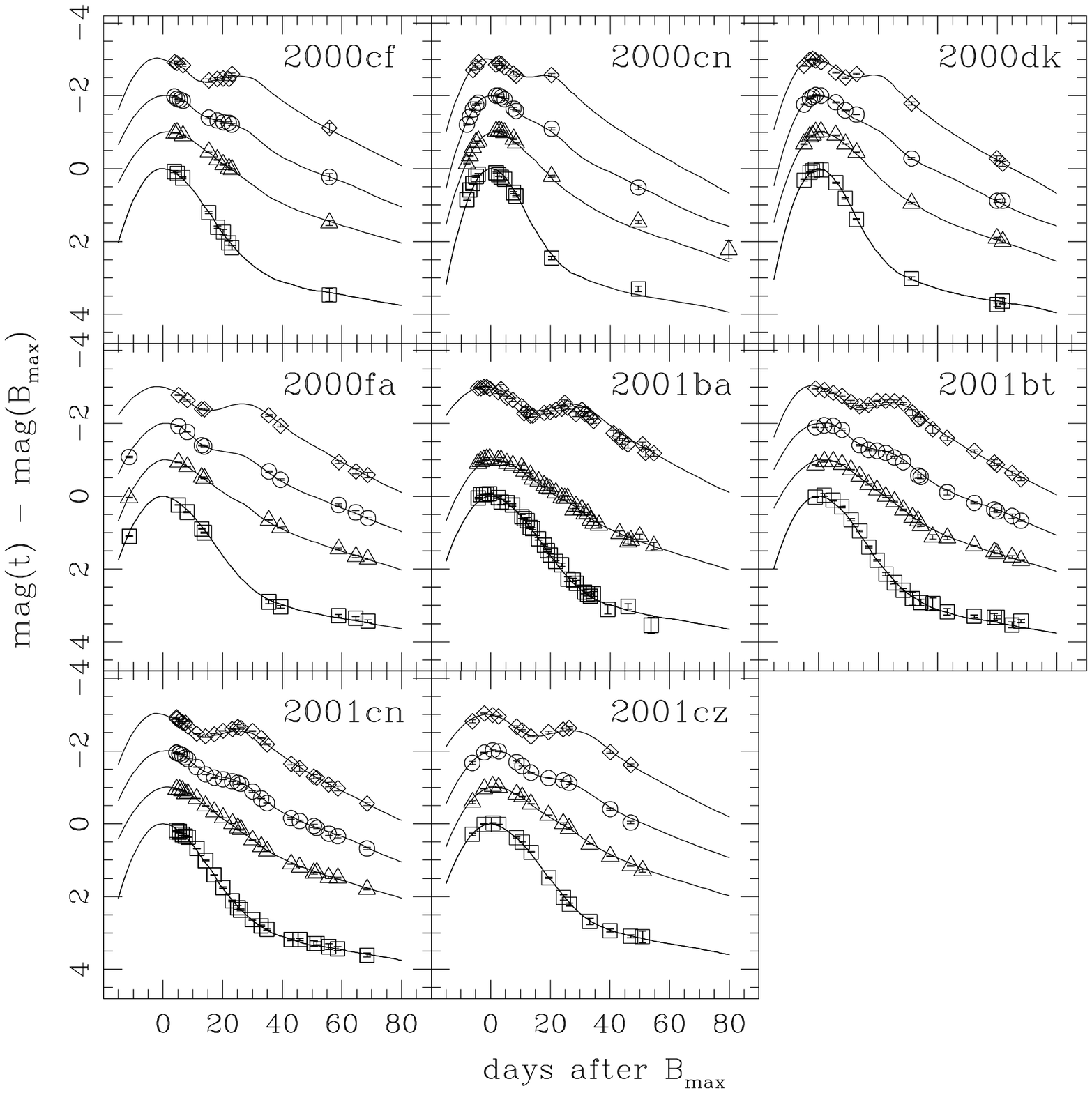}
\caption{Results of the fits to the multicolor light curves. The
different filters (symbols) are: $B$ (squares), $V$ (triangles), $R$
(circles) and $I$ (rhombuses).  To avoid overlap between different
filters we applied shifts to the relative magnitudes: $\Delta B$,
$\Delta V$-1, $\Delta R$-2, $\Delta I$-3. \label{fig_16}}
\end{figure}

With the relations for $M_{max}(\Delta \rm{m_{15}})$ in different
filters (RHS of Equation \ref{eq:4}), we can now apply our fitting
technique minimizing the $\chi^{2}$ function in Equation \ref{eq:5} to
obtain the distance modulus of a large sample of SNe in the Hubble
flow. We use the complete sample of 89 SNe in Table \ref{table_2} with $z>0.01$
and $\Delta \rm{m_{15}}\leq 1.70$ mag. The results of the fits are
presented in Table \ref{table_5} where we give the best fit parameters
($\Delta {\rm m_{15}}, \, E(B-V)_{host}, \,\mu_{0}$) including their
statistical errors from the covariance matrix. We plot all the light
curves and best fit models in Figures \ref{fig_7}-\ref{fig_16}. We
obtained a median value of $\chi^{2}_{\nu}=1.3$ for all the fits.

\begin{figure}
\plotone{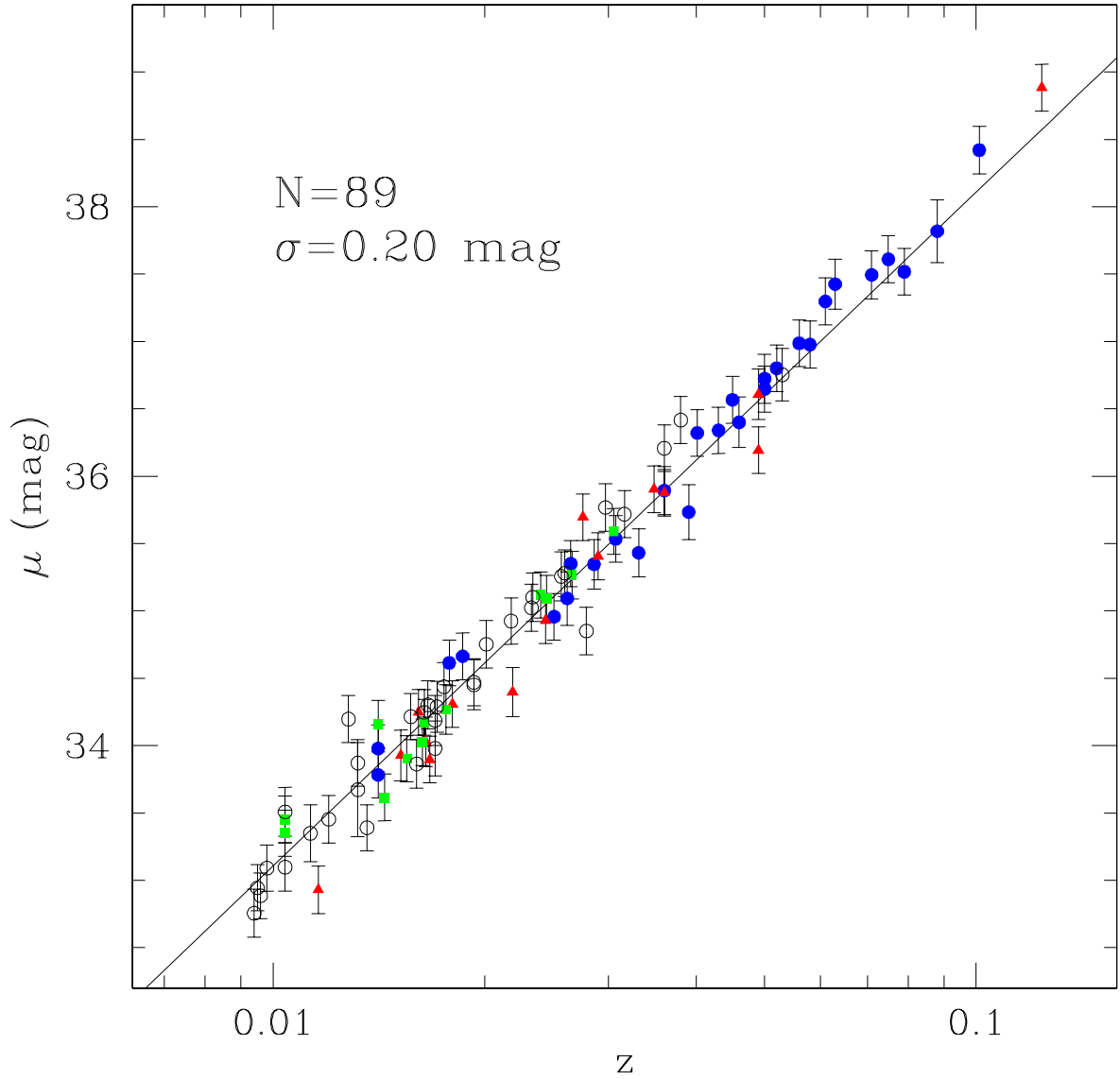}
\caption{Hubble Diagram of the complete sample of SNe with $z>0.01$
and $\Delta m_{15} \leq 1.70$ mag (89). The symbols correspond to different
subsamples of SNe: Cal\'an/Tololo (filled circles), CfAI (filled
triangles), Krisciunas et al. (filled squares), and CfAII (open
circles). \label{fig_6}}
\end{figure}

In Figure \ref{fig_6} we plot the Hubble diagram constructed with
results of the fits to the 89 SNe in the Hubble flow.  The final error
in the distance modulus is obtained by adding in quadrature the
statistical uncertainties from the covariance matrix and 0.17 mag,
which is the maximum natural dispersion in the relations
$M_{max}(\Delta \rm{m_{15}})$ obtained in Section \ref{sec:3}.  At
these low redshifts, we can ignore the cosmological effects and fit
the Hubble law as the second order expansion of the luminosity
distance with $\Omega_{M}=0$ and $\Omega_{\Lambda}=0$:
$d_{L}(z)\simeq (cz/H_{0})(1 + z/2)$.  We find that the dispersion in the
Hubble diagram is $\sigma=0.20$ mag using 89 SNe. This is a $9\%$ error in distance
to a single SN, a very accurate distance estimator considering the
large sample of SNe studied. As a comparison, \citet{Jha:2006b} obtains 
$\sigma=0.18$ mag using 95 SNe in the same redshift range. 

In Table \ref{table_6}, we compare the dispersion in the Hubble
diagram with other results in the literature, using the same SNe as in
the original papers. In general we find similar or smaller values for the
dispersion, with the Cal\'an/Tololo subsample giving the fit with the
lowest dispersion. The dispersion varies in the range
$\sigma=0.14-0.21$ mag with our technique and between $\sigma =
0.14-0.24$ mag for results in the literature, using identical samples
of SNe.  This further demonstrates that our technique performs equally
well or better than other existing techniques in the literature. To
reduce the dispersion in the Hubble diagram further will require new
uniform data sets, realistic priors that allow different reddening
laws for the host galaxies, and careful attention to the calibration
of the telescope/filter/detector system required by the S-corrections.

\section{Conclusions}

We have presented an empirical method to fit multicolor light curves
of Type Ia supernovae and estimate their luminosity distances. This
technique combines what we think are the advantages from two widely
used methods in the literature: the \mbox{$\Delta$m$_{15}$} template
fitting method \citep{Phillips:1999}, and MLCS \citep{Riess:1996}.

Our basic fitting algorithm uses a set of 14 $BVRI$ light curve
templates with different values of \mbox{$\Delta$m$_{15}(B)$} and a
simple triangle weighting function. A linear combination of the
templates, weighted by this function, is introduced directly into the
$\chi^{2}$ fitting, avoiding the construction of a secondary grid of
interpolated templates. This allows us to add or change templates
trivially, and also allows us to introduce other parameters to be fit
in the $\chi^{2}$ minimization. The $\chi^{2}$ fit returns the
$1\sigma$ uncertainties in the best fit parameters from the covariance
matrix.

From a sample of 94 nearby Type Ia SNe ($z\lesssim 0.1$) we
established linear relations between the absolute magnitudes at
maximum $M_{max}$ and \mbox{$\Delta$m$_{15}$} in $BVRI$ filters. These
relations are valid in the range $0.80 < \Delta {\rm m}_{15} {\rm
[mag]} < 1.70$ and they are consistent with other results presented in
the literature \citep{Phillips:1999, Garnavich:2004, Germany:2004}. We
studied the relations using different subsamples of SNe, associated
with different surveys and publications in the literature:
Cal\'an/Tololo, CfAI, Krisciunas et al., CfAII and a nearby sample
($z\lesssim 0.01$) with distances obtained from SBF method. The
results from different subsamples are consistent in almost all the
cases when we compare the slopes and zero-points of the linear relations in
the same filter, but disturbing differences do exist mainly in $B$ and
$V$ absolute magnitudes at $\Delta m_{15} = 1.1$.  Further progress
will require new light curve data sets, realistic priors that allow
different reddening laws, and careful attention to the calibration of
the telescope/filter/detector system required by the S-corrections.

We have constructed a Hubble diagram with low-$z$ SNe fitting the
light curves of 89 SNe in the Hubble flow. The dispersion of the
Hubble diagram is $\sigma=0.20$ mag or an error of $\sim 9\%$ in
distance to single objects, consistent with other fitting techniques
\citep{Jha:2006b, Germany:2004}.  We compared the dispersion in the
Hubble diagram using our technique with other results in the
literature. In general our technique give similar or smaller
dispersions than published results when the comparison is made using
identical samples of SNe.

\acknowledgments

We thank A.~Clocchiatti, C.~Contreras, K.~Krisciunas, M.~Phillips and
B.~Schmidt for helpful discussions and comments.  J.~L.~P. thanks the
staff of CTIO for all their support during his stay in La Serena,
where this paper was partially written.

We made use of the NASA/IPAC Extragalactic Database (NED), which is
operated by the Jet Propulsion Laboratory, California Institute of
Technology, under contract with NASA. J.~L.~P. was supported in part
by an OSU Astronomy Department Fellowship. This work was supported by
the Space Telescope Science Institute through grant HST GO-9860, and
by the U.~S. National Science Foundation under grant AST-0206329.


\clearpage


\input{tab1}

\clearpage
\input{tab2}

\clearpage
\input{tab3}

\clearpage
\input{tab4}

\clearpage
\input{tab5}

\clearpage
\input{tab6}

\clearpage

\clearpage

\clearpage

\clearpage 

\clearpage 

\clearpage

\clearpage

\end{document}

%% file: tab1.tex
\begin{deluxetable}{lcccccc}
\tablewidth{0pt}
\tabletypesize{\small}
\tablecaption{Basic information of the light curve templates in the template set. \label{table_1}}
\tablehead{
\colhead{SN}         & \colhead{\mbox{$\Delta$m$_{15}$}}  & \colhead{$E(B-V)_{Gal}$} & \colhead{$\Delta$m$^{V}$}      &
\colhead{$\Delta$m$^{R}$}          & \colhead{$\Delta$m$^{I}$}  &
\colhead{Ref} \\
(1)   &  (2)  & (3)  &  (4) & (5) & (6) & (7) \\
}
\startdata
1991T   &  0.94 & 0.022  &  -0.02  &  -0.01  &   0.00  & 1, 2 \\
1991bg  &  1.93 & 0.040  &  -0.05  &  -0.14  &  -0.10  & 1, 5, 6  \\
1992A   &  1.47 & 0.017  &  -0.03  &  -0.03  &  -0.05  & 1, 4 \\
1992al  &  1.11 & 0.034  &   0.00  &  -0.02  &  -0.01  & 1, 3 \\
1992bc  &  0.87 & 0.022  &   0.00  &   0.00  &  -0.02  & 1, 3 \\
1992bo  &  1.69 & 0.027  &   0.00  &   0.00  &  -0.02  & 1, 3 \\
1994D   &  1.34 & 0.022  &   0.00  &   0.00  &  -0.10  & 7 \\
1994ae  &  0.94 & 0.031  &  -0.01  &   0.00  &   0.00  & 8 \\
1995D   &  1.05 & 0.058  &   0.00  &   0.00  &   0.00  & 8 \\
1995al  &  0.89 & 0.014  &  -0.01  &  -0.01  &  -0.01  & 8 \\
1996X   &  1.25 & 0.069  &  -0.02  &   0.00  &  -0.07  & 9 \\
1998bu  &  1.05 & 0.025  &   0.00  &  -0.01  &  -0.05  & 10, 11 \\
1999aa  &  0.83 & 0.015  &  -0.02  &  -0.02  &  -0.03  & 12, 13 \\
2001el  &  1.12 & 0.014  &  -0.03  &  -0.01  &  -0.06  & 14 \\
\enddata
\tablecomments{The columns are: (1) name of the supernova; (2) \mbox{$\Delta$m$_{15}(B)$} from the  
$B$ light curve template; (3) $V_{max}-V_{t_{0}}$; (4) $R_{max}-R_{t_{0}}$; (5) $I_{max}-I_{t_{0}}$;
(6) References containing the data with the light curves used to construct the templates.}
\tablerefs{
(1) \citealt{Hamuy:1996d}; (2) \citealt{Lira:1998}; (3) \citealt{Hamuy:1996c}; 
(4) Suntzeff et al. (unpublished); (5) \citealt{Fili:1992};
(6) \citealt{Leib:1993}; (7) Smith et al. (unpublished); (8) \citealt{Riess:1999a};
(9) Covarrubias et al. (unpublished); (10) \citealt{Jha:1999}; (11) \citealt{Suntzeff:1999};
(12) \citealt{KK:2000}; (13) \citealt{Jha:2006a}; (14) \citealt{KK:2003}}
\end{deluxetable}

%% file: tab2.tex


\begin{deluxetable}{llc}

\tablewidth{0pt}
\tabletypesize{\small}
\tablecaption{Different samples of SNe used to study the relations between absolutes magnitudes at maximum ($M_{max}$) and $\Delta {\rm m}_{15}$. \label{table_2}}
\tablehead{
\colhead{Sample}  &  \colhead{SN} & \colhead{Ref} \\
}
\startdata
Cal\'an/Tololo        &   90O, 90T, 90Y, 90af, 91S, 91U, 91ag, 92ae, 92al,  &  1 \\ 
                      &   92bc, 92J, 92P, 92ag, 92aq, 92au, 92bg, 92bh,  &   \\ 
                      &   92bk, 92bl, 92bo, 92bp, 92br, 92bs, 93B, 93O,  &    \\ 
                      &   93ag, 93ah   & \\ 
                      &   & \\
CfAI                 &   93ac, 93ae, 94M, 94Q, 94S, 94T, 95E, 95ac, 95ak, &  2 \\ 
                      &   95bd, 96C, 96ab, 96bl, 96bo, 96bv   &   \\
                      &   & \\
Krisciunas et al.     &   99aa, 99cp, 99dk, 99ee, 99ek, 99gp, 00bh, 00bk, &  3, 4, 5, 6, 7\\
                      &   00ca, 00ce, 01ba, 01bt, 01cn, 01cz  & \\
                      &   & \\
CfAII                &   97E, 97Y, 97cw, 97dg, 97do, 98D, 98V, 98ab, 98co,  &  8 \\
                      &   98dk, 98dx, 98ec, 98ef, 98eg, 99X, 99aa, 99ac,  &  \\
                      &   99cc, 99cw, 99dq, 99ef, 99ej, 99ek, 99gd, 99gp, &  \\ 
                      &   00B, 00ce, 00cf, 00cn, 00dk, 00fa   &  \\
                      &   & \\
Nearby                &   72E, 80N, 81B, 86G, 89B, 90N, 91T, 92A, 94D, 95D,  &   1, 2, 7, 8, 11, 12, 13, 14, 15, 16,\\
                      &   96X, 96bk, 98aq, 98bu, 99by, 02bo  &  17, 18, 19, 20, 21, 22, 23, 24, 25  \\
                      &       &  \\
                      &       &   \\
\enddata
\tablecomments{The columns are: (1) Name of the data sample; (2) Supernova in the sample; (3) References.}
\tablerefs{
(1) \citealt{Hamuy:1996c}; (2) \citealt{Riess:1999a}; (3) \citealt{KK:2000}; 
(4) \citealt{KK:2001}; (5) \citealt{Stritzinger:2002};
(6) \citealt{KK:2004b}; (7) \citealt{KK:2004c}; (8) \citealt{Jha:2006a}; (9) \citealt{Ardeberg:1973};
(10) \citealt{van Genderen:1975}; (11) \citealt{Lee:1972}; (12) \citealt{Cousins:1972}; (13) \citealt{Hamuy:1991};
(14) \citealt{Barbon:1982}; (15) \citealt{Buta:1983}; (16) \citealt{Phillips:1987}; (17) \citealt{Cristiani:1992};
(18) \citealt{Schaefer:1987}; (19) \citealt{Barbon:1990}; (20) \citealt{Wells:1994}; (21) \citealt{Lira:1998}; 
(22) Smith et al. (unpublished); (23) Covarrubias et al. (unpublished);
(24) Boffi et al. (unpublished); (25) \citealt{Garnavich:2004}.}
\end{deluxetable}


%% file: tab3.tex
\begin{deluxetable}{cccccc}
\tablewidth{0pt}
\tabletypesize{\small}
\tablecaption{Best fit linear relations to $M_{max}$ versus $\Delta m_{15}$ in $BVRI$ filters for the complete sample (ALL). The assumed linear relations are of the form: $M_{max} = a + b(\Delta m_{15} -1.1)$ in the range $0.80 \leq \Delta m_{15}\leq 1.70$. We also give the linear fits using a cut in host galaxy color excess at $E(B-V)_{host}=0.06$ mag. \label{table_3}} 
\tablehead{
\colhead{Filter}  & \colhead{$a$}  & \colhead{$b$} & \colhead{$\sigma$} & \colhead{$\chi^{2}_{\nu}$} & \colhead{N} \\ 
}
\startdata
\cutinhead{ALL}
$B$ &  $-$19.319(022) & 0.634(077) & 0.17 & 0.73 & 94 \\
$V$ &  $-$19.246(019) & 0.606(069) & 0.14 & 0.68 & 94 \\
$R$ &  $-$19.248(025) & 0.566(101) & 0.13 & 0.45 & 73 \\
$I$ &  $-$18.981(020) & 0.524(079) & 0.15 & 0.90 & 87 \\
\cutinhead{$E(B-V)_{host} \leq 0.06$ mag}
$B$ &  $-$19.325(024) & 0.636(082) & 0.16 & 0.91 & 62 \\
$V$ &  $-$19.247(022) & 0.598(075) & 0.14 & 0.85 & 62 \\
$R$ &  $-$19.251(031) & 0.522(012) & 0.14 & 0.55 & 44 \\
$I$ &  $-$18.975(023) & 0.464(090) & 0.15 & 1.07 & 57 \\
\cutinhead{$E(B-V)_{host} > 0.06$ mag}
$B$ &  $-$19.286(051) & 0.753(255) & 0.19 & 0.39 & 32 \\
$V$ &  $-$19.233(044) & 0.748(226) & 0.15 & 0.36 & 32 \\
$R$ &  $-$19.229(044) & 0.761(216) & 0.12 & 0.28 & 29 \\
$I$ &  $-$18.983(038) & 0.825(196) & 0.15 & 0.52 & 30 \\
\enddata

\tablecomments{The columns are: (1) filter; (2) zero-point of the linear fits with uncertainties in 
parentheses, in units of 0.001 mag; (3) slope of the linear fits with uncertainties in parentheses, in units of 0.001 mag; 
(4) rms scatter of each sample around the linear fits; (5) $\chi^{2}$ per degree of freedom of the fits; 
(6) number of SNe.}

\end{deluxetable}

%% file: tab4.tex
\begin{deluxetable}{cccccc}
\tablewidth{0pt}
\tabletypesize{\small}
\tablecaption{Best fit linear relations $M_{max}$ versus $\Delta m_{15}$ in $BVRI$ filters for different sub-samples of Table \ref{table_2}. \label{table_4}} 
\tablehead{
\colhead{Filter}  & \colhead{$a$}  & \colhead{$b$} & \colhead{$\sigma$} & \colhead{$\chi^{2}_{\nu}$} & \colhead{N} \\ 
}
\startdata
\cutinhead{C\'alan/Tololo}
$B$ & $-$19.376(037) & 0.814(118) & 0.11 & 0.80 & 27 \\
$V$ & $-$19.295(032) & 0.773(105) & 0.11 & 0.97 & 27 \\
$R$ & $-$19.249(062) & 1.003(364) & 0.13 & 0.56 & 9 \\
$I$ & $-$18.974(032) & 0.559(130) & 0.14 & 1.49 &  23 \\
\cutinhead{CfAI}
$B$ & $-$19.224(055) & 0.880(210) &  0.19 & 0.56 &  15  \\
$V$ & $-$19.167(049) & 0.659(187) &  0.16 & 0.61 &  15  \\
$R$ & $-$19.269(059) & 0.721(205) &  0.11 & 0.33 &  14  \\
$I$ & $-$19.049(053) & 0.787(193) &  0.16 & 0.98 &  14  \\
\cutinhead{Krisciunas et al.}
$B$ & $-$19.279(084) & 1.224(570) &  0.21 & 0.51 & 13  \\
$V$ & $-$19.218(082) & 1.159(551) &  0.16 & 0.32 & 13  \\
$R$ & $-$19.192(091) & 0.868(546) &  0.12 & 0.21 & 12   \\
$I$ & $-$18.958(078) & 0.406(520) &  0.12 & 0.21 & 13   \\
\cutinhead{CfAII}
$B$ & $-$19.209(049) & 0.246(153) &  0.15 & 0.46 & 30 \\
$V$ & $-$19.171(046) & 0.307(145) &  0.14 & 0.40 & 30 \\
$R$ & $-$19.188(044) & 0.349(142) &  0.14 & 0.47 & 30 \\
$I$ & $-$18.941(042) & 0.283(141) &  0.17 & 0.77 & 30 \\
\cutinhead{Nearby}
$B$ & $-$19.343(051) & 0.166(284) & 0.14 & 0.42 &  16 \\
$V$ & $-$19.265(046) & 0.398(262) & 0.12 & 0.34 &  16 \\
$R$ & $-$19.259(046) & 0.515(263) & 0.12 & 0.37 &  15 \\
$I$ & $-$18.986(044) & 0.478(242) & 0.13 & 0.47 &  14 \\
\enddata

\tablecomments{The columns are: (1) filter; (2) zeropoint of the linear fits with uncertainties in 
parentheses, in units of 0.001 mag; (3) slope of the linear fits with uncertainties in parentheses, in units of 0.001 mag; 
(4) rms scatter of each subsample/filter around the linear fits; (5) $\chi^{2}$ per degree of freedom of the linear fits; 
(6) number of SNe.}
\end{deluxetable}

%% file: tab5.tex
\begin{deluxetable}{lcccc}
\tablewidth{0pt}
\tabletypesize{\small}
\tablecaption{Final results of the \mbox{$\Delta$m$_{15}$} fits to the light curves of 89 Type Ia SNe in the Hubble flow \mbox{($0.01 < z < 0.10$)}. \label{table_5}}
\tablehead{
\colhead{SN}         & \colhead{$z$}  & \colhead{$\mu_{0}(\sigma)$} & 
\colhead{$E(B-V)_{host}(\sigma)$}          & \colhead{$\Delta$m$_{15}(\sigma)$} \\
\,  (1)   &  (2)  & (3)  &  (4) & (5) 
}
\startdata
1990O & 0.0307 & 35.535(0.028) & 0.003(0.008) & 1.019(0.017) \\ 
1990T & 0.0401 & 36.321(0.026) & 0.017(0.010) & 1.120(0.017)  \\ 
1990Y & 0.0390 & 35.733(0.112) & 0.194(0.023) & 1.273(0.030)  \\ 
1990af & 0.0500 & 36.648(0.017) & 0.006(0.006) & 1.681(0.010)  \\ 
1991S & 0.0560 & 36.987(0.023) & 0.001(0.005) & 1.036(0.021)  \\ 
1991U & 0.0331 & 35.431(0.049) & 0.092(0.022) & 0.929(0.018)  \\ 
1991ag & 0.0141 & 33.782(0.027) & 0.000(0.004) & 0.975(0.023) \\ 
1992J & 0.0460 & 36.400(0.079) & 0.031(0.030) & 1.662(0.045)  \\ 
1992P & 0.0265 & 35.350(0.021) & 0.006(0.006) & 1.088(0.019)  \\ 
1992ae & 0.0750 & 37.610(0.041) & 0.007(0.011) & 1.212(0.025) \\ 
1992ag & 0.0262 & 35.092(0.103) & 0.106(0.027) & 1.162(0.026) \\ 
1992al & 0.0141 & 33.978(0.040) & 0.000(0.005) & 0.916(0.050) \\ 
1992aq & 0.1010 & 38.421(0.042) & 0.000(0.009) & 1.273(0.022) \\ 
1992au & 0.0610 & 37.297(0.040) & 0.005(0.011) & 1.259(0.021) \\ 
1992bc & 0.0186 & 34.663(0.032) & 0.000(0.007) & 0.838(0.030) \\ 
1992bg & 0.0360 & 35.893(0.057) & 0.002(0.011) & 1.179(0.020) \\ 
1992bh & 0.0450 & 36.567(0.033) & 0.085(0.011) & 1.065(0.018) \\ 
1992bk & 0.0580 & 36.977(0.040) & 0.018(0.015) & 1.677(0.010) \\ 
1992bl & 0.0430 & 36.340(0.025) & 0.000(0.005) & 1.515(0.030) \\ 
1992bo & 0.0178 & 34.614(0.014) & 0.015(0.006) & 1.688(0.001) \\ 
1992bp & 0.0790 & 37.518(0.024) & 0.000(0.006) & 1.238(0.018) \\ 
1992br & 0.0880 & 37.819(0.158) & 0.075(0.053) & 1.690(0.010) \\ 
1992bs & 0.0630 & 37.426(0.072) & 0.033(0.025) & 1.076(0.017) \\ 
1993B & 0.0710 & 37.495(0.052) & 0.037(0.019) & 1.097(0.014) \\ 
1993H & 0.0251 & 34.957(0.034) & 0.153(0.012) & 1.700(0.010) \\ 
1993O & 0.0520 & 36.801(0.036) & 0.016(0.013) & 1.207(0.012) \\ 
1993ac & 0.0490 & 36.608(0.077) & 0.134(0.032) & 1.068(0.021) \\ 
1993ae & 0.0180 & 34.308(0.026) & 0.000(0.004) & 1.560(0.025) \\ 
1993ag & 0.0500 & 36.724(0.063) & 0.064(0.015) & 1.225(0.017) \\ 
1993ah & 0.0286 & 35.346(0.067) & 0.028(0.030) & 1.285(0.017) \\ 
1994M & 0.0244 & 34.930(0.029) & 0.113(0.011) & 1.376(0.022) \\ 
1994Q & 0.0290 & 35.405(0.041) & 0.037(0.018) & 1.064(0.012)  \\ 
1994S & 0.0161 & 34.247(0.014) & 0.000(0.006) & 0.859(0.010)  \\ 
1994T & 0.0360 & 35.877(0.036) & 0.111(0.013) & 1.475(0.010)  \\ 
1995E & 0.0116 & 32.929(0.045) & 0.720(0.013) & 1.170(0.019) \\
1995E & 0.0116 & 33.141(0.055) & 0.721(0.020) & 0.830(0.030)  \\ 
1995ac & 0.0490 & 36.193(0.035) & 0.090(0.012) & 0.830(0.010)  \\ 
1995ak & 0.0219 & 34.398(0.066) & 0.208(0.021) & 1.303(0.017)  \\ 
1995bd & 0.0152 & 33.927(0.081) & 0.140(0.032) & 0.909(0.035)  \\ 
1996C & 0.0276 & 35.695(0.030) & 0.041(0.009) & 1.011(0.013)  \\ 
1996ab & 0.1240 & 38.884(0.036) & 0.000(0.009) & 1.064(0.015)  \\ 
1996bl & 0.0348 & 35.903(0.038) & 0.061(0.015) & 0.830(0.030)  \\ 
1996bo & 0.0165 & 34.028(0.072) & 0.305(0.028) & 0.830(0.010)  \\ 
1996bv & 0.0167 & 33.897(0.036) & 0.190(0.014) & 0.940(0.019)  \\ 
1997E & 0.0132 & 33.871(0.032) & 0.054(0.012) & 1.514(0.024)  \\ 
1997Y & 0.0166 & 34.302(0.053) & 0.065(0.015) & 1.132(0.026)  \\ 
1997bp & 0.0094 & 32.756(0.052) & 0.132(0.017) & 0.830(0.010) \\ 
1997bq & 0.0095 & 32.944(0.018) & 0.086(0.006) & 1.467(0.010)  \\ 
1997cn & 0.0175 & 34.437(0.061) & 0.192(0.023) & 1.686(0.010)  \\ 
1997cw & 0.0160 & 33.863(0.060) & 0.322(0.009) & 0.906(0.060)  \\ 
1997dg & 0.0297 & 35.768(0.053) & 0.085(0.015) & 1.203(0.024)  \\ 
1997do & 0.0104 & 33.508(0.064) & 0.081(0.022) & 0.830(0.020)  \\ 
1998D  & 0.0132 & 33.672(0.305) & 0.186(0.060) &  1.697(0.010) \\  
1998V & 0.0170 & 34.187(0.075) & 0.064(0.023) & 0.983(0.040)  \\ 
1998ab & 0.0279 & 34.851(0.043) & 0.212(0.013) & 0.832(0.010)  \\ 
1998bp & 0.0104 & 33.097(0.053) & 0.251(0.020) & 1.700(0.010)  \\ 
1998co & 0.0171 & 34.288(0.084) & 0.172(0.031) & 0.845(0.014)  \\ 
1998dk & 0.0120 & 33.453(0.051) & 0.140(0.017) & 1.140(0.022)  \\ 
1998dx & 0.0530 & 36.753(0.096) & 0.050(0.032) & 1.063(0.028)  \\ 
1998ec & 0.0201 & 34.752(0.048) & 0.161(0.022) & 1.074(0.014)  \\ 
1998ef & 0.0170 & 33.979(0.112) & 0.132(0.036) & 0.830(0.010)  \\ 
1998eg & 0.0234 & 35.102(0.060) & 0.085(0.021) & 1.062(0.035)  \\ 
1998es & 0.0096 & 32.885(0.016) & 0.076(0.006) & 0.917(0.007)  \\ 
1999X & 0.0257 & 35.256(0.062) & 0.079(0.029) & 0.958(0.027)  \\ 
1999aa & 0.0157 & 34.215(0.023) & 0.000(0.007) & 0.837(0.020)  \\ 
1999ac & 0.0098 & 33.090(0.026) & 0.081(0.009) & 1.067(0.010)  \\ 
1999cc & 0.0316 & 35.718(0.037) & 0.035(0.012) & 1.408(0.026)  \\ 
1999cp & 0.0104 & 33.452(0.034) & 0.002(0.007) & 0.838(0.030)  \\ 
1999cw & 0.0113 & 33.349(0.124) & 0.010(0.042) & 0.830(0.020)  \\ 
1999dk & 0.0141 & 34.157(0.056) & 0.000(0.020) & 0.830(0.010)  \\ 
1999dq & 0.0136 & 33.390(0.016) & 0.092(0.006) & 0.941(0.010)  \\ 
1999ee & 0.0104 & 33.349(0.015) & 0.230(0.005) & 0.830(0.010)  \\ 
1999ef & 0.0380 & 36.417(0.044) & 0.000(0.010) & 1.065(0.015)  \\ 
1999ej & 0.0128 & 34.197(0.043) & 0.051(0.017) & 1.534(0.029)  \\ 
1999ek & 0.0176 & 34.268(0.070) & 0.157(0.027) & 1.058(0.034)  \\ 
1999gd & 0.0193 & 34.451(0.074) & 0.411(0.022) &  1.180(0.025) \\  
1999gp & 0.0260 & 35.280(0.032) & 0.034(0.010) & 0.832(0.010)  \\ 
2000B & 0.0193 & 34.470(0.038) & 0.075(0.015) & 1.300(0.016)  \\ 
2000bh & 0.0240 & 35.118(0.021) & 0.001(0.003) & 1.129(0.009)  \\ 
2000bk & 0.0266 & 35.267(0.048) & 0.146(0.020) & 1.700(0.010)  \\ 
2000ca & 0.0245 & 35.092(0.017) & 0.000(0.005) & 0.858(0.010)  \\ 
2000ce & 0.0164 & 34.170(0.037) & 0.513(0.016) & 0.998(0.031)  \\  
2000cf & 0.0360 & 36.207(0.031) & 0.023(0.013) & 1.157(0.019)  \\ 
2000cn & 0.0233 & 35.023(0.036) & 0.099(0.013) & 1.700(0.010)  \\ 
2000dk & 0.0164 & 34.246(0.013) & 0.004(0.006) & 1.690(0.010)  \\ 
2000fa & 0.0218 & 34.925(0.029) & 0.063(0.011) & 1.049(0.016)  \\ 
2001ba & 0.0305 & 35.590(0.013) & 0.000(0.004) & 1.054(0.010)  \\ 
2001bt & 0.0144 & 33.616(0.036) & 0.214(0.011) & 1.170(0.014)  \\ 
2001cn & 0.0155 & 33.901(0.019) & 0.128(0.006) & 1.152(0.010)  \\ 
2001cz & 0.0163 & 34.025(0.032) & 0.096(0.009) & 0.977(0.018)  \\ 
\enddata
\tablecomments{The columns are: (1) name of the supernova; (2) redshift in the 
CMB frame; (3) best fit distance modulus ($h=0.72$), statistical errors in parentheses;
(4) best fit color excess $E(B-V)_{host}$, statistical errors in parentheses; 
(5) best fit $\Delta {\rm m}_{15}$, statistical errors in parentheses.}

\end{deluxetable}

%% file: tab6.tex
\begin{deluxetable}{lccc}
\tablewidth{0pt}
\tabletypesize{\small}
\tablecaption{Comparison of rms dispersion in the Hubble diagram using the technique presented in this paper with other results in the literature. \label{table_6}}
\tablehead{
\colhead{Literature} & \colhead{$\sigma_{\rm Literature}$}  & \colhead{$\sigma_{\rm This\, Work}$} & \colhead{Number of SNe}  
}

\startdata
Phillips et al. (1999)  & 0.14  & 0.14 & 26 \\
Knop et al. (2003)      & 0.20  & 0.16 & 23 \\
Riess et al. (2004)     & 0.24  & 0.19 & 68 \\
Germany et al. (2004)   & 0.23  & 0.19 & 42 \\
Reindl et al. (2005)    & 0.21  & 0.21 & 71 \\
\enddata
\end{deluxetable}

%% file: pres.bbl
\begin{thebibliography}{}

\expandafter\ifx\csname natexlab\endcsname\relax\def\natexlab#1{#1}\fi

\bibitem[Ajhar et~al.(2001)]{Ajhar:2001}
Ajhar, E.~A., Tonry, J.~T., Blakeslee, J.~P., Riess, A.~G., \& Schmidt, B.~P. 2001, \apj, 559, 584

\bibitem[Altavilla et~al.(2004)]{Altavilla:2004}
Altavilla, G., Fiorentino, G., Marconi, M., et~al. 2004, \mnras, 349, 1344

\bibitem[Ardeberg \& de Groot(1973)]{Ardeberg:1973}
Ardeberg, A., \& de Groot, M. 1973, \aap, 28, 295

\bibitem[Arnett(1982)]{Arnett:1982}
Arnett, W.~D. 1982, \apj, 253, 785 

\bibitem[Baker, Davis \& Lin(2000)]{Baker:2000}
Baker, J.~E., Davis, M., \& Lin, H. 2000, \apj, 536, 112

\bibitem[Barbon, Ciatti \& Rosino(1982)]{Barbon:1982}
Barbon, R., Ciatti, F., \& Rosino, L. 1982, \aap, 116, 35

\bibitem[Barbon et~al.(1990)]{Barbon:1990}
Barbon, R., Bennetti, S., Rosino, L., et al. 1990, \aap, 237, 79

\bibitem[Bevington \& Robinson(1990)]{Bevington}
Bevington, P.~R., \& Robinson, D.~K. 1992, \textit{Data Reduction and Error Analysis for the Physical Sciences}, McGraw-Hill, Inc.   

\bibitem[Branch(1987)]{Branch:1987}
Branch, D. 1987, \apjl, 316, L81

\bibitem[Buta \& Turner(1983)]{Buta:1983}
Buta, R.~J., \& Turner, A. 1983, \pasp, 95, 72

\bibitem[Cardelli et~al.(1989)]{Cardelli:1989}
Cardelli, J.~A., Clayton, G.~C., \& Mathis, J.~S. 1989, \apj, 345, 245

\bibitem[Carroll, Press, \& Turner(1992)]{Carroll:1992}
Carroll, S.~M., Press, W.~H., \& Turner, E.~L. 1992, \araa, 30, 499

\bibitem[Cousins(1972)]{Cousins:1972}
Cousins, A,~W.,~J. 1972, BVS, 700, 1

\bibitem[Cristiani et~al.(1992)]{Cristiani:1992}
Cristiani, S., Capellaro, E., Turatto, M., et~al. 1992, \aap, 259, 63

\bibitem[Filippenko et~al.(1992)]{Fili:1992}
Filippenko, A.~V., Richmond, M.~W., Branch, D., et~al. 1992, \aj, 104, 1543 

\bibitem[Garnavich et~al.(2004)]{Garnavich:2004}
Garnavich, P.~M., Bonanos, A.~Z., Krisciunas, K., et~al. 2004, \apj, 613, 1120 

\bibitem[Germany et~al.(2004)]{Germany:2004}
Germany, L.~M., Reiss, D.~J., Schmidt, B.~P., Stubbs, C.~W., \& Suntzeff, N.~B. 2004, \aap, 415, 863

\bibitem[Goldhaber et~al.(2001)]{Goldhaber:2001}
Goldhaber, G., Groom, D.~E., Kim, A., et~al. 2001, \apj, 558, 359

\bibitem[Hawkins et~al.(2003)]{Hawkins:2003}
Hawkins, E., Maddox, S., Cole, S., et~al. 2003, \mnras, 346, 78

\bibitem[Hamuy et~al.(1991)]{Hamuy:1991}
Hamuy, M., Phillips, M.~M., Maza, J., et~al. 1991, \aj, 102, 208

\bibitem[Hamuy et~al.(1993)]{Hamuy:1993}
Hamuy, M., Phillips, M.~M., Wells, L.~A., \& Maza, J. 1993, \pasp, 105, 787

\bibitem[Hamuy et~al.(1996{\natexlab{a}})]{Hamuy:1996a}
Hamuy, M., Phillips, M.~M., Suntzeff, N.~B., Schommer, R.~A., Maza, J., \& Aviles, R. 1996{\natexlab{a}}, \aj, 112, 2391

\bibitem[Hamuy et~al.(1996{\natexlab{b}})]{Hamuy:1996b}
Hamuy, M., Phillips, M.~M., Suntzeff, N.~B., Schommer, R.~A., Maza, J., \& Aviles, R. 1996{\natexlab{b}}, \aj, 112, 2398

\bibitem[Hamuy et~al.(1996{\natexlab{c}})]{Hamuy:1996c}
Hamuy, M., Phillips, M.~M., Suntzeff, N.~B., et~al. 1996{\natexlab{c}}, \aj, 112, 2408

\bibitem[Hamuy et~al.(1996{\natexlab{d}})]{Hamuy:1996d}
Hamuy, M., Phillips, M.~M., Suntzeff, N.~B., Schommer, R.~A., Maza, J., Smith, R.~C., Lira, P., \& Aviles, R. 1996{\natexlab{d}}, 
\aj, 112, 2438

\bibitem[Jha et~al.(2006{\natexlab{a}})]{Jha:2006a}
Jha, S., Kirshner, R.~P., Challis, P., et~al. 2006, AJ, accepted

\bibitem[Jha, Riess, \& Kirshner (2006{\natexlab{b}})]{Jha:2006b}
Jha, S., Riess, A.~G., \&  Kirshner, R.~P. 2006, ApJ, submitted

\bibitem[Jha et~al.(1999)]{Jha:1999}
Jha, S., Garnavich, P.~M., Kirshner, R.~P., et~al. 1999, \apjs, 125, 73 

\bibitem[Kim, Goobar, \& Perlmutter (1996)]{Kim:1996}
Kim, A., Goobar, A., \& Perlmutter, S. 1996, \pasp, 108, 190

\bibitem[Krisciunas et~al.(2000)]{KK:2000}
Krisciunas, K., Hastings, N.~C., Loomis, K., et~al. 2000, \apj, 539, 658

\bibitem[Krisciunas et~al.(2001)]{KK:2001}
Krisciunas, K., Phillips, M.~M., Stubbs, C., et~al. 2001, \aj, 122, 1616

\bibitem[Krisciunas et~al.(2003)]{KK:2003}
Krisciunas, K., Suntzeff, N.~B., Candia, P., et~al. 2003, \aj, 125, 166

\bibitem[Krisciunas et~al.(2004{\natexlab{b}})]{KK:2004b}
Krisciunas, K., Phillips, M.~M., Suntzeff, N.~B., et~al. 2004{\natexlab{b}}, \aj, 127, 1664

\bibitem[Krisciunas et~al.(2004{\natexlab{c}})]{KK:2004c}
Krisciunas, K., Suntzeff, N.~B., Phillips, M.~M., et~al. 2004{\natexlab{c}}, \aj, 128, 303

\bibitem[Landy(2002)]{Landy:2002}
Landy, S.~D. 2002, \apjl, 567, L1  

\bibitem[Lee et~al.(1972)]{Lee:1972}
Lee, T.~A., Wamsteker, W., Wisniewski, W.~Z., \& Wdowiak, T.~J. 1972, \apjl, 177, L59  

\bibitem[Leibundgut(1988)]{Leib:1988}
Leibundgut, B. 1988, Ph.D.~Thesis, University of Basel

\bibitem[Leibundgut et~al.(1993)]{Leib:1993}
Leibundgut, B., Kirshner, R.~P., Phillips, M.~M. 1993, \aj, 105, 301

\bibitem[Leibundgut(2000)]{Leib:2000}
Leibundgut, B. 2000, \aapr, 10, 179 

\bibitem[Lira(1995)]{Lira:1995}
Lira, P. 1995, Masters thesis, Universidad de Chile

\bibitem[Lira et~al.(1998)]{Lira:1998}
Lira, P., Hamuy, M., Wells, L.~A., et~al. 1998, \aj, 115, 234

\bibitem[Marzke et~al.(1995)]{Marzke:1995}
Marzke, R.~O., Geller, M.~J., da Costa, L.~N., \& Huchra, J.~P. 1995, \aj, 110, 447

\bibitem[Nugent, Kim, \& Perlmutter(2002)]{Nugent:2002}
Nugent, P., Kim, A., \& Perlmutter, S. 2002, \pasp, 114, 803

\bibitem[Parodi et~al.(2000)]{Parodi:2000} Parodi, B.~R., Saha, A., 
Sandage, A., \& Tammann, G.~A.\ 2000, \apj, 540, 634 

\bibitem[Perlmutter et~al.(1997)]{Perlmutter:1997}
Perlmutter, S., Gabi, S., Goldhaber, G., et~al. 1997, \apj, 483, 565

\bibitem[Perlmutter et~al.(1999)]{Perlmutter:1999}
Perlmutter, S., Aldering, G., Goldhaber, et~al. 1999, \apj, 517, 565

\bibitem[Phillips et~al.(1987)]{Phillips:1987}
Phillips, M.~M., Phillips, A.~C., Heathcote, S.~R., et~al. 1987, \pasp, 99, 592

\bibitem[Phillips(1993)]{Phillips:1993}
Phillips, M.~M. 1993, \apjl, 413, L105

\bibitem[Phillips {et~al.}(1999)]{Phillips:1999}
Phillips, M.~M., Lira, P., Suntzeff, N.~B., Schommer, R.~A., Hamuy, M., \& Maza, J. 1999, \aj, 118, 1766

\bibitem[Press et~al.(1988)]{Press:1988}
Press, W.~H., Teukolsky, S.~A., Vetterling, W.~T., \& Flannery, B.~P. 1988, \textit{Numerical Recipes in C}, Cambridge University Press. 

\bibitem[Pskovskii(1977)]{Pskovskii:1977}
Pskovskii, I.~P. 1977, Soviet Astronomy, 21, 675

\bibitem[Riendl et~al.(2005)]{Riendl:2005}
Riendl, B., Tammann, G.~A., Sandage, A., \& Saha A. 2005, \apj, 624, 532 

\bibitem[Riess, Press, \& Kirshner(1996)]{Riess:1996}
Riess, A.~G., Press, W.~H., \& Kirshner, R.~P. 1996, \apj, 473, 88

\bibitem[Riess et~al.(1998)]{Riess:1998}
Riess, A.~G., Filippenko, A.~V., Challis, P., et~al. 1998, \aj, 116, 1009

\bibitem[Riess et~al.(1999{\natexlab{a}})]{Riess:1999a}
Riess, A.~G., Kirshner, R.~P., Schmidt, B.~P., et~al. 1999{\natexlab{a}}, \aj, 117, 707

\bibitem[Riess et~al.(1999{\natexlab{b}})]{Riess:1999b}
Riess, A.~G., Filipenko, A.~V., Li, W., \& Schmidt, B.~P. 1999{\natexlab{b}}, \aj, 118, 2668  

\bibitem[Riess et~al.(2004)]{Riess:2004}
Riess, A.~G., Strolger, L.~G., Tonry, J., et~al. 2004, \apj, 607, 665

\bibitem[Schaefer(1987)]{Schaefer:1987}
Schaefer, B.~E. 1987, \apjl, 323, L47 

\bibitem[Schmidt et~al.(1998)]{Schmidt:1998}
Schmidt, B.~P., Suntzeff, N.~B., Phillips M.~M., et~al. 1998, \apj, 507, 46

\bibitem[Schlegel, Finkbeiner, \& Davis (1998)]{Schlegel:1998}
Schlegel, D.~J., Finkbeiner, D.~P., \& Davis, M.~J. 1998, \apj, 500, 525

\bibitem[Stritzinger et~al.(2002)]{Stritzinger:2002}
Stritzinger, M., Hamuy, M., Suntzeff N.~B., et~al. 2002, \aj, 124, 2100

\bibitem[Suntzeff et~al.(1999)]{Suntzeff:1999}
Suntzeff, N.~B., Phillips, M.~M., Covarrubias, R., et~al. 1999, \aj, 117, 1175

\bibitem[Suntzeff(2000)]{Suntzeff:2000}
Suntzeff, N.~B. 2000, in AIP Conf. Proc. 522, Cosmic Explosions, ed. S.~S. Holt \& W.~W. Zhang (Melville, NY: AIP), 65

\bibitem[Tripp(1998)]{Tripp:1998}
Tripp, R. 1998, \aap, 331, 815

\bibitem[Tripp \& Branch(1999)]{Tripp:1999}
Tripp, R. \& Branch, D. 1999, \apj, 525, 209

\bibitem[Tonry et~al.(2003)]{Tonry:2003}
Tonry, J.~L., Schmidt, B.~P., Barris, B., et~al. 2003, \apj, 594, 1

\bibitem[van Genderen(1975)]{van Genderen:1975}
van Genderen, A.~M. 1975, \aap, 45, 429

\bibitem[Wang et~al.(2003)]{Wang:2003}
Wang, L., Goldhaber, G., Aldering, G., \& Perlmutter, S. 2003, \apj, 590, 994

\bibitem[Wang et~al.(2005)]{Wang:2005}
Wang, X., Wang, L., Xu, Z., et~al. 2005, \apj, 620, 87

\bibitem[Wells et~al.(1994)]{Wells:1994}
Wells, L.~A., Phillips, M.~M., Suntzeff, N.~B., et~al. 1994, \aj, 108, 2233

\bibitem[Zehavi et~al.(1998)]{Zehavi:1998}
Zehavi, I., Riess, A.~G., Kirshner, R.~P., \& Dekel, A. 1998, \apj, 503, 483

\bibitem[Zehavi et~al.(2002)]{Zehavi:2002}
Zehavi, I., Blanton, M.~R., Frieman, J.~A., et~al. 2002, \apj, 571, 172

\end{thebibliography}
